\documentclass[pra,twocolumn,superscriptaddress,amsmath,amssymb,longbibliography]{revtex4-1}

\usepackage{graphicx}
\usepackage{color}

\mathchardef\mhyphen="2D 

\usepackage[colorlinks=true,linkcolor=red,citecolor=blue,urlcolor=blue]{hyperref}

\begin{document}

\title{Unconventional Chemical Bonding of Lanthanide-OH Molecules}

\author{Jacek K{\l}os}
\affiliation{Department of Physics, Temple University, Philadelphia, Pennsylvania 19122, USA}
\author{Eite Tiesinga} 
\affiliation{Joint Quantum Institute, National Institute of Standards and Technology and University of Maryland, Gaithersburg,  Maryland 20899, USA}
\author{Lan Cheng}
\affiliation{Department of Chemistry, The Johns Hopkins University, Baltimore, Maryland 21218, USA}
\author{Svetlana Kotochigova}
\email{skotoch@temple.edu}
\address{Department of Physics, Temple University, Philadelphia, Pennsylvania 19122, USA}

\date{\today}

\begin{abstract}
We present a theoretical study of the low lying adiabatic relativistic electronic states of  lanthanide monohydroxide  (Ln-OH)
molecules near their linear equilibrium geometries. In particular, we focus on  heavy, magnetic DyOH and ErOH 
relevant to fundamental symmetry tests. We use a  restricted-active-space self-consistent field method combined with spin-orbit coupling as well as a  relativistic coupled-cluster method to determine  
ground and excited electronic states.  In addition, electric dipole and magnetic moments are computed with the self-consistent field method.
Analysis of the results from both methods shows that the dominant molecular configuration of the ground state is one where 
an electron from the partially filled and submerged 4f orbital of the lanthanide atom moves to the hydroxyl group, leaving 
the closed outer-most 6s$^2$ lone electron pair of the lanthanide atom intact in sharp contrast to the bonding in alkaline-earth monohydroxides and YbOH, where an electron from the outer-most s shell moves to the hydroxyl group.
For  linear molecules the projection of the total electron angular momentum on the symmetry axis is a conserved quantity with quantum number $\Omega$ and we study the polynomial $\Omega$ dependence of the energies of the ground states as well as their electric and magnetic moments. We find that for both molecules $\Omega$  lies between $-15/2$ and $+15/2$, where the degenerate states with the lowest energy have $|\Omega|=15/2$ and 1/2 for  DyOH and ErOH, respectively. The zero field splittings among these $\Omega$ states is approximately $hc\times 1\,000$~cm$^{-1}$, where $h$ is the Planck constant and $c$ is the speed of light in vacuum. We find that the permanent dipole moments for both triatomics are fairly small at 0.23 atomic units and are mostly independent of $\Omega$. The magnetic moments are closely related to that of the corresponding atomic Ln$^+$ ion in an excited electronic state. From the polynomial $\Omega$ dependences, we also realize that the  total electron angular momentum is to good approximation conserved and has a quantum number of 15/2 for both triatomic molecules. We describe how this observation can be used to construct effective Hamiltonians containing  spin-spin operators.
\end{abstract}

\maketitle

\section{Introduction}

Lanthanide atoms have a submerged open 4f electron-shell lying underneath a 6s$^2$-closed shell. Experimental breakthroughs in the creation of ultracold quantum gases of these atoms with their large magnetic moments \cite{Lu2011, Lu2012, Frisch2012, Aikawa2012} have opened a  scientific playground in which to study quantum magnetism at the interface between condensed matter and atomic physics. In addition, these magnetic lanthanides have a large electronic orbital angular momentum leading to anisotropies, {\it i.e.} orientation dependencies, in their mutual interactions. Anisotropic interactions are crucial for using ultracold lanthanide atoms in spin-based quantum computing and  simulations of orbitronics \cite{Go2017, Bernevig2005}.
 
The idea of integrating  distinct physical components into a hybrid quantum system has also received significant attention. Hybrid approaches promise to take advantage of each components best properties, thereby realizing new tools for quantum information processing, condensed matter physics, and high-precision measurements.
Our  objective in this paper is to explore a novel class hybrid quantum system, hybrid magnetic lanthanide molecules, where lanthanide (Ln) atoms are brought and bound together with hydroxyl radical. 
These triatomic molecules are expected to possess  features that are not present in lanthanide atoms. 

Heavy atoms bound to hydroxyl or alcoxy groups  stand out as candidates for precision measurements in search of  physics beyond the standard model \cite{Isaev2017, Kozyryev2017, Hutzler2019, Hutzler2020, BAugenbraun2020}. Among these systems are molecules containing heavy atoms with octupole deformed nuclei, such as isotopes of Fr and Ra,  but also  lanthanides and even actinides \cite{Chen2024}.  
Octupole deformed nuclei have a  Schiff moment and thus are sensitive probes of  violations of the charge-parity (CP) symmetry  \cite{Auerbach1996, Dobaczewski2005, Skripnikov2020, Yu2021, Flambaum2022}. In these molecules, the electronic structure can enhance the effect of the  nuclear Schiff moment  and  lead to  observable parity breaking terms in the ro-vibrational molecular Hamiltonian. These systems can also enhance the effects of a nuclear electric dipole moment  \cite{Sushkov1984, Flambaum1994, Flambaum2014, Flambaum2022}.

A second reason for our interest in DyOH and ErOH is that they might be amenable 
to laser cooling. Laser cooling of atoms is possible for atoms that contain a non-degenerate two level system, where the higher energy state predominantly decays back to the lower energy state by the emission of a ``spontaneous'' photon, typically in or around the optical frequency domain. Such closed systems naturally occur in alkali-metal and alkaline-earth atoms. On the other hand, Dy and Er have a complex level structure, often with metastable states, and successful cooling seems a challenging proposition. Nevertheless,  Refs.~\cite{McClelland2006,Frisch2012,Lu2010}  showed that despite the possible optical leaks, the large electronic angular momenta of  Er and Dy states make laser cooling feasible. 
Recently, laser cooling has also been reported for CaOH, SrOH, and YbOH \cite{Vilas2024, Kozyryev2017, BAugenbraun2020a} even though molecules have even more ways to have optical leaks. For these three molecules cooling is possible due to the existence of so-called vertical transitions, where the lower- and higher-energy potential energy curves have very similar equilibrium geometries and harmonic spring constants or vibrational structure.

Previous studies  of the electronic ground state of LnOH used density functional theory (DFT)  \cite{Harb2019} and predicted that the equilibrium geometry of  these molecules for all atoms in the lanthanide series is linear with the oxygen atom in the middle. Moreover, the lanthanide atom loses one electron from its 6s shell to OH, the bond between the lanthanide and the hydroxide is a covalent triple bond, and the partially filled 4f shell is only weakly involved in  the  bond. 

We will describe calculations of the relativistic electronic eigenstates  of DyOH and ErOH using
restricted-active-space self-consistent-field calculations combined with computed spin-orbit matrix elements within the OpenMolcas package \cite{OpenMolcas} as well as using a relativistic coupled-cluster method within the CFOUR package \cite{cfour,Rel.Method}.  Within the self-consistent-field method,  we  also compute electric and magnetic dipole moments.  We obtain a different bonding model between the lanthanide atom and the hydroxyl group than that found within DFT.  The molecules are still  linear  at their equilibrium geometries and  the lanthanide atom still loses one electron to OH.  We, however, find that the 4f shell participates significantly in the bonding.  We show that the 4f$^n$ configuration of the lanthanide atom with $n=10$ and 12 for Dy and Er, respectively, mixes with the 2p$^5$ configuration of the tightly-bound ground-state OH,
leading to ``spin-orbit'' states of the  $4{\rm f}^{n-1}6{\rm s}^2+2{\rm p}^6$ configuration as the ground
and lowest-energy  eigenstates. The spin-orbit components of the $4{\rm f}^{n}6{\rm s}+2{\rm p}^6$
configuration, where the 6s lanthanide orbital has lost an electron, have significantly higher energies. The authors of
Ref.~\cite{Yamamoto2015}  found a similar behavior in the isoelectronic DyF using a four-component
relativistic configuration-interaction approach. Recently, Ref.~\cite{Chen2024} confirmed this bond for DyF.

\section{Results}
\subsection{The DyOH and ErOH bond and zero field splittings}

\begin{figure*}
\includegraphics[scale=0.3,trim=0 0 0 0,clip]{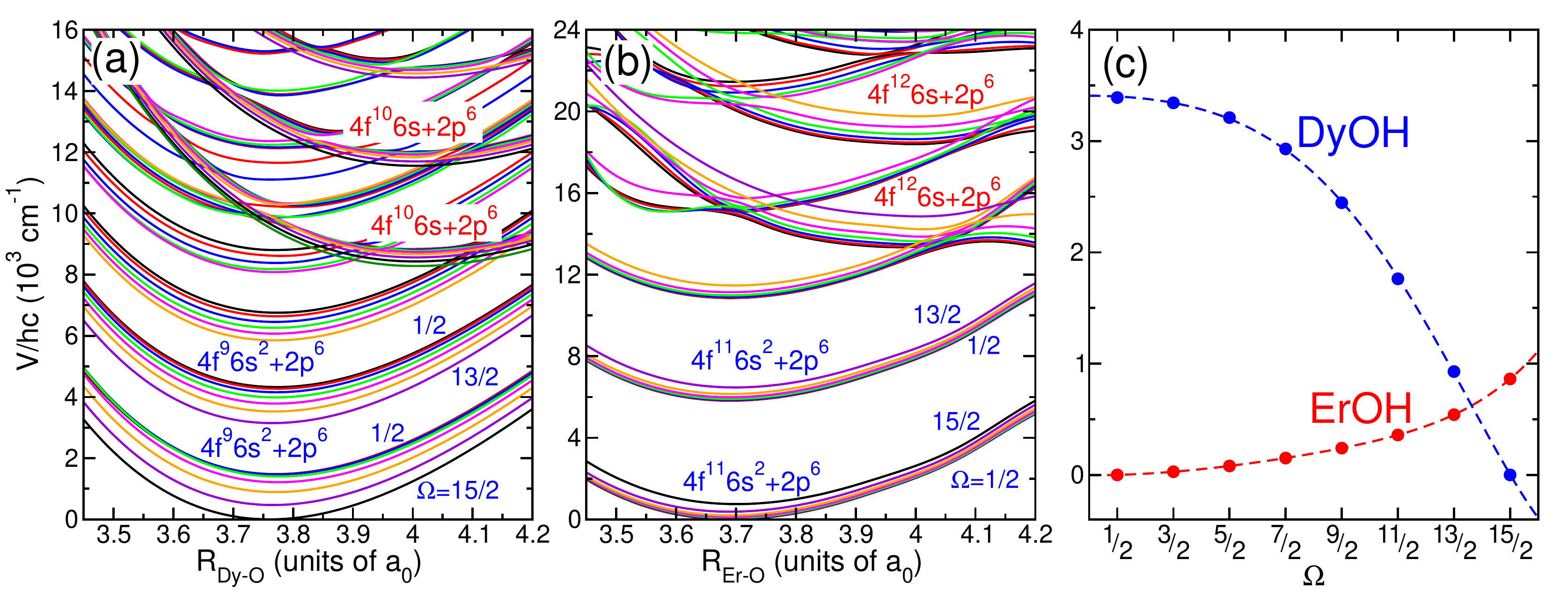}
\caption{Potential energies of the $4{\rm f}^{n-1}6{\rm s}^2+2{\rm p}^6$ and $4{\rm f}^{n}6{\rm s}+2{\rm p}^6$ electronic configurations of DyOH (panel (a)) and ErOH (panel (b)) near their  equilibrium geometries as functions of the $X$-O separation with $X={\rm Dy}$ or Er for a linear geometry and at a fixed O-H separation of $1.80a_0$.
For each configuration the curves correspond to states with different $\Omega$ (or more precisely $|\Omega|$).
Curves with the same color have the same $|\Omega|$.
The zero of energy is at the equilibrium geometry of energetically lowest potential.
The potentials  have been obtained with self-consistent-field 
calculations using  basis sets that {\it do not} include excitations into 6p and 5d molecular orbitals. 
Panel (c) shows the splittings among the $\Omega$ states of the $4{\rm f}^{n-1}6{\rm s}^2+2{\rm p}^6$ configuration (colored circles). The energies have been obtained with self-consistent-field 
calculations using basis sets that {\it do} include excitations into 6p and 5d molecular orbitals. 
The dashed curves through the markers are  fits to these data and are described in the text.}
\label{fig:LnOH}
\end{figure*}

We begin by making some general observations about our triatomic systems.
Heavy-element-containing molecular systems are difficult to model as relativistic electronic-structure methods
are required. Further complicating the calculations for triatomic molecules is that the (relativistic) Born-Oppenheimer (BO) approximation can more 
easily break down than in diatomic molecules. Hence, a large number of adiabatic potential energy surfaces and 
electronic states are
involved in the atomic motion. A theoretical analysis of the large number of electronic states in terms of spherical spin
tensor operators, describing the coupling of the spins and angular momenta in the system, is convenient in
order to account for non-adiabatic coupling among the states. For our DyOH and ErOH systems with their odd
number of electrons, the relevant states turn out to correspond to different orientations of the combined spins and
orbital angular momenta of the electrons in the submerged 4f shell.

The DyOH  and ErOH triatomics at their equilibrium geometries are linear molecules, where the lanthanide atom is bound
to the O-side of the OH radical. Consequently, the natural body-fixed Cartesian coordinate system  is one
where the positive $z$ axis points from the lanthanide to the oxygen atom and the $x$ and $y$ axes form a plane perpendicular  to this symmetry axis. For a non-relativistic description of linear molecules, electronic eigenstates  
are  labeled by the conserved total electronic spin ${\bf S}$ and  projection quantum number $\Lambda$ of the total electronic  orbital angular momentum ${\bf L}$ along our body-fixed $z$ axis. In a relativistic description of the electrons,
molecular electronic levels  only conserve $\Omega=\Lambda+\Sigma$, where $\Sigma$ is the  projection
quantum number of ${\bf S}$ along the $z$ axis. 

For linear geometries, potentials $V_{c\Omega}$ and kets $|c,\Omega\rangle$ will denote electronic eigenenergies and eigenstates, respectively. Here, $c$ labels the dominant molecular configuration of an eigenstate. 
For example, $c=4{\rm f}^{10}6{\rm s}+2{\rm p}^6$, $4{\rm f}^96{\rm s}^2+2{\rm p}^6$, 
or $4{\rm f}^96{\rm s}6{\rm p}+2{\rm p}^6$ for DyOH. 
Finally,  electronic eigenstates with quantum numbers $-\Omega$ and $+\Omega$ for $\Omega>0$ are degenerate.  
For  half-integer $\Omega$, these degenerate pairs are Kramers' doublets \cite{Kramers1930}.

After these preliminaries, we are  ready to describe our results.
Figures~\ref{fig:LnOH}(a) and (b) show one-dimensional slices through the three-dimensional relativistic potentials $V_{c\Omega}$ obtained with non-relativistic restricted-active-space self-consistent-field (RAS-SCF) calculations that do not include excitations to the 6p and 5d molecular orbitals at each molecular geometry, followed by the evaluation of spin-orbit matrix elements
among a limited number of non-relativistic eigenstates and diagonalization of the electronic Hamiltonian within this subset of states.
In the absence of the 6p and 5d orbitals, the SCF simulations correspond to complete-active-space SCF calculations (CAS-SCF) with 4f and 6s orbitals. 

In Figs.~\ref{fig:LnOH}(a) and (b),  we draw the potentials for linear molecules as functions of the Ln-O separation near the equilibrium geometry of DyOH and ErOH with the O-H separation fixed at its equilibrium separation of $1.80a_0$ for both triatomics.   All separations are in units of the Bohr radius $a_0$  and energies in units of $hc$, where $h$ is the Planck constant and $c$
is the speed of light in vacuum. Values for these constants in SI units as well as that for the elementary charge $e$
and the Bohr magneton $\mu_{\rm B}$, both used later on, are taken from Ref.~\cite{CODATA2018}. We provide more details about the electronic structure calculations in Methods.

First we note  that  the relativistic potentials in Figs.~\ref{fig:LnOH}a) and b)   are
assigned to the $4{\rm f}^{n-1}6{\rm s}^2+2{\rm p}^6$ or $4{\rm f}^{n}6{\rm s}+2{\rm p}^6$
configurations. In fact, several ``bundles'' of levels, well separated in energy, are assigned with these two configurations. 
Crucially, we predict that many states assigned to the $4{\rm f}^{n-1}6{\rm s}^2+2{\rm p}^6$ configuration that have a lower energy than those of the $4{\rm f}^{n}6{\rm s}+2{\rm p}^6$ configuration.  The largest energy difference is approximately $hc\times 10\, 000$ cm$^{-1}$. 
 As we will show later on in this paper adding the virtual 6p and 5d orbitals in  the SCF calculations significantly  increases the level density and leads to strong mixing of the 6s, 6p, and 5d molecular orbitals and makes  assignment by dominant molecular configurations for excited states difficult. Still, we find that these differences in basis sets do not change our conclusion regarding the energy ordering of configurations. Of course, calculations including the 6p and 5d orbitals are more accurate.

Second, we observe that for DyOH in Fig.~\ref{fig:LnOH}(a)  the absolute ground-state potential  has quantum number ${|\Omega|=15/2}$  with a minimum energy  when the Dy-O and O-H separations are $3.76a_0$ and  $1.80a_0$, respectively.  This compares well with our relativistic coupled-cluster calculations with single, double, and perturbative triple excitations (CCSD(T)) for this state. We find that the potential minimum occurs at a Dy-O separation of $3.60a_0$, which is slightly shorter than that found with the SCF simulations due to a better inclusion of dynamic electronic correlation. The minimum energy of the potentials for states with $|\Omega|<15/2$ of the lowest-energy DyOH  ``bundle'' increases with decreasing $|\Omega|$ corresponding to an inverted spin-orbit interaction. There are  $2\times8=16$  levels in this grouping accounting for the twofold degeneracies. The second lowest ``bundle'' of states also assigned with the $4{\rm f}^{n-1}6{\rm s}^2+2{\rm p}^6$ configuration has $2\times 7=14$ states and contains pairs of levels for $|\Omega|=13/2$ down  to 1/2 in order of increasing energy. Finally, the bundles of potential energy curves assigned with the $4{\rm f}^{n-1}6{\rm s}^2+2{\rm p}^6$ configuration
are parallel within the accuracy of our simulations, {\it i.e.} have nearly the  same equilibrium separation and force or spring constant (and thus harmonic frequency).

For ErOH in Fig.~\ref{fig:LnOH}(b) the absolute ground state are the ${|\Omega|=1/2}$ leveld of the $4{\rm f}^{n-1}6{\rm s}^2+2{\rm p}^6$ configuration.  The corresponding potential has its minimum at Er-O and  O-H separations of $3.69a_0$ and $1.80a_0$, respectively.  The minimum energy of the potentials for states with $|\Omega|>1/2$ of this lowest ``bundle'' increases with increasing $|\Omega|$ up to $|\Omega|=15/2$.  The second lowest ``bundle'' has $2\times 7=14$ states and contains pairs of levels for $|\Omega|=1/2$ up to 13/2 in order of increasing energy.
The potential energy curves assigned with the $4{\rm f}^{n-1}6{\rm s}^2+2{\rm p}^6$ configuration
are again parallel to good approximation.

For completeness, we note that from the SCF calculations, we find that the total electron spin of the non-relativistic DyOH and ErOH ground state  are $S=5/2$ and 3/2, respectively, corresponding to the spin-stretched state for both triatomics.
Finally, the equilibrium O-H separation in both triatomic molecules is only slightly smaller than the
spectroscopically determined value for the equilibrium separation of the ground state of the free OH radical
at $1.8324a_0$. 

Figure ~\ref{fig:LnOH}(c) shows the DyOH and ErOH potential energies of the energetically lowest sixteen relativistic eigenstates, {\it i.e.} the first bundle or group of levels,
at their equilibrium geometry as functions of $|\Omega|$. These  potential energies are evaluated from RAS-SCF calculations that include excitations to the 6p and 5d molecular orbitals. As mentioned before, this group of levels still remains  energetically well separated from other groups of levels. Clearly shown is the inverted spin-orbit splittings for DyOH and the ``normal'' one for ErOH.
The zero of energies are chosen to be at the $|\Omega|=15/2$ and 1/2 states for DyOH and ErOH, respectively.

\begin{figure*}
\includegraphics[scale=0.27,trim=0 0 0 0,clip]{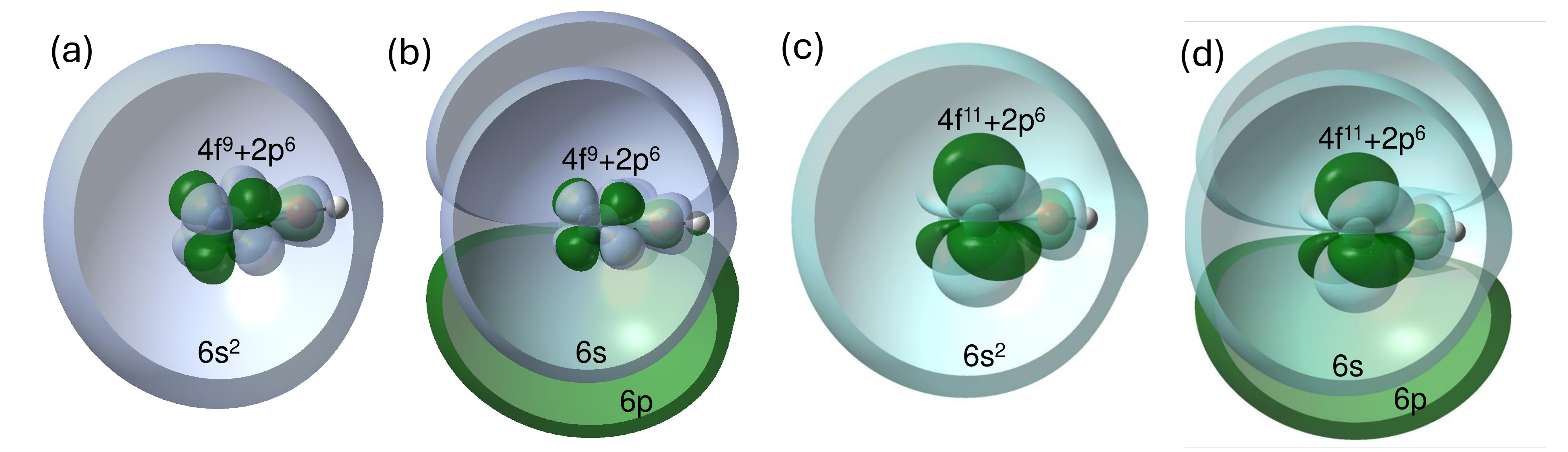}
\caption{Isosurfaces of electronic Kohn-Sham molecular orbitals  (MOs) for DyOH (panels (a) and (b)) and ErOH (panels (c) and (d)) at their equilibrium, linear geometries. In all panels the small, nearly hidden cyan, red, and gray balls correspond to the locations of the lanthanide, oxygen, and hydrogen atom, respectively.
Panels (a) and (c)  show some of the MOs for the $4{\rm f}^{n-1}6{\rm s}^2+2{\rm p}^6$ configuration where
the 4f and 2p orbitals overlap. The chosen 4f orbitals for DyOH and ErOH resemble the f$_{xyz}$  and f$_{yz^2}$ cubic or tesseral harmonic, respectively, while the 2p orbital resembles the p$_z$ cubic harmonic with the lobes aligned along the $z$ or Ln-O axis. Similarly, panels (c) and (d) show some of the MOs for the $4{\rm f}^{n-1}6{\rm s}6{\rm p}+2{\rm p}^6$ configuration, where now the largest feature corresponds to the 6p orbital resembling the p$_x$ or $p_y$ cubic harmonic. }
\label{fig:combined} 
\end{figure*}

The equilibrium potential energies or zero-field splittings (ZFSs) of the $c=4{\rm f}^96{\rm s}^2+2{\rm p}^6$  ground-state configuration of DyOH in Fig.~\ref{fig:LnOH}(c) and with our choice for the zero of energy are well described by polynomial
\begin{eqnarray}
V_{c\Omega}/(hc\ {\rm cm}^{-1}) &=& 
  3\,407.59	-25.6016 \,\Omega^2 -1.305\,06\, \Omega^4 \nonumber\\
	&&\quad  + \,0.012\,168\, \Omega^6
	\label{eq:energyDyOH}
\end{eqnarray}
for $|\Omega|=1/2,\cdots,15/2$, which ensures that eigenstates with $-\Omega$ and $\Omega$ are degenerate.
Similarly, for the  $c=4{\rm f}^{11}6{\rm s}^2+2{\rm p}^6$ ground-state configuration of ErOH, the zero-field splittings  are well described by
\begin{eqnarray}
  V_{c\Omega}/(hc\ {\rm cm}^{-1}) &=&
 -3.65105+14.5863\, \Omega^2 -0.204\,103 \,\Omega^4\nonumber\\
	&&\quad  + \,0.003\,885\,86\,\Omega^6 \,.
	\label{eq:energyErOH}
  \end{eqnarray}
As the potentials for linear DyOH and ErOH are nearly parallel with Ln-O separation, the coefficients of the terms proportional to $\Omega^{2n}$ with $n>0$  weakly depend on this separation. 
In fact, most separation dependence is isolated in the coefficient of the $\Omega^0$ term.

The energies of the levels in Eqs.~(\ref{eq:energyDyOH}) and  (\ref{eq:energyErOH}) can be equivalently described by
a spherical tensor-operator expansion once we 
assume that the electronic wavefunction is also an eigenstate of the square of the total electronic angular momentum of
the molecule, ${\bf j}^2$.  Here, we use the observations that the group of levels contain one state for each $|\Omega|$ up 
to 15/2, that the  $2{\rm p}^6$ configuration of the OH$^{-}$ negative ion has a closed and thus spin-less electron shell, and 
that the lowest energy state of the  $4{\rm f}^{n-1}6{\rm s}^2$ configuration of the atomic Dy$^+$ and Er$^+$ ion has total {\it atomic} angular momentum $j_{\rm atom}=15/2$ solely due to the open 4f shell \cite{Kramida2023}.
Hence, we are justified in assuming that  the total {\it molecular} electron angular momentum ${\bf j}$ has quantum number  $j=15/2$.

In a laboratory-fixed coordinate system $(X,Y,Z)$ we can then write effective Hermitian potential operator
\begin{eqnarray}
  \hat V &=& 
a_0+a_2 T_{2}(j,j)\cdot C_2(\hat R) \label{eq:BOtensor}\\
 && \phantom{a_0} +a_4T_{4}(T_2(j,j),T_2(j,j))\cdot C_4(\hat R)\nonumber \\
 && \phantom{a_0}+a_6T_{6}(T_3(j,T_2(j,j)),T_3(j,T_2(j,j)))\cdot C_6(\hat R)
 \nonumber
\end{eqnarray}
with the connection $|c,\Omega\rangle \to |j\Omega\rangle$, energy coefficients $a_i$ and dot product $R_k\cdot S_k=\sum_{q=-k}^k(-1)^q R_{kq}S_{k-q}$ for arbitrary spherical tensor operators $R_{kq}$ and 
$S_{kq}$ of  rank $k$, a non-negative integer. 
Moreover, tensor operators $T_{kq}(\cdot,\cdot)$ of  rank $k=2,3,\cdots$ are constructed from rank-1 total electronic angular momentum operator ${\bf j}$ 
and, finally, $C_{kq}(\hat R)$ are spherical harmonic functions, where unit vector $\hat R=(\theta,\varphi)$ describes the orientation of the symmetry axis of the linear triatomic molecule in  the laboratory-fixed coordinate system.  
We follow the definition and notation for constructing tensor operators of Ref.~\cite{Brink1993}, functions $C_{kq}(\hat R)$ satisfy $C_{kq}(\hat 0)=\delta_{q0}$,
and quantity $j$ in tensor operators, such as $T_2(j,j)$ and $T_3(j,T_2(j,j))$, is an abbreviation for dimensionless operator ${\bf j}/\hbar$, where $\hbar$ is the reduced Planck constant, the atomic unit of angular momentum.

For the relevant eigenstates $|c\Omega\rangle$ and, more precisely, $|j\Omega\rangle$ in the body-fixed coordinate system with the $z$ axis along
the symmetry axis of the linear triatomic molecule and thus $\hat R=\hat 0$, the operator $\hat V$ in Eq.~(\ref{eq:BOtensor}) leads to eigen energies
\begin{eqnarray}
V_{c\Omega}&=&a_0+a_2(3\Omega^2-j(j+1))/\sqrt{6} \label{eq:Vc} \\
  && \qquad {}+ a_4 \langle j 4 \,\Omega 0| j\Omega\rangle\times \langle j || T_4(.,.)||j\rangle
  \nonumber\\
  && \quad\qquad{}+ a_6 \langle j 6\, \Omega 0| j\Omega\rangle\times \langle j || T_6(.,.)||j\rangle
  \nonumber
\end{eqnarray}
using the Wigner-Eckart theorem with Clebsch-Gordan coefficients $\langle j_1 j_2 m_1 m_2| jm\rangle$ 
and reduced matrix elements $\langle j || T_k(.,.)||j\rangle$ \cite{Brink1993}.
We have explicitly evaluated the matrix element for the operator multiplying $a_2$.
The reduced matrix elements multiplying $a_4$ and $a_6$ can be evaluated with
repeated use of Eq.~(5.5) of Ref.~\cite{Brink1993}.
We also realize that the Clebsch-Gordan coefficients $\langle j k \,\Omega 0| j\Omega\rangle$ with $k=2,4,6,\cdots$ in Eq.~(\ref{eq:Vc}) are  polynomials in $\Omega^2$ with degree $k/2$.
For example, $\langle j 4 \,\Omega 0| j\Omega\rangle$ is proportional to
$35 \Omega^4+5[5-6j(j+1)]\Omega^2+3(j+2)(j+1)j(j-1)$.
Values for  $a_{i}$ with even $i=0,2,\cdots$ can found from a comparison with  Eqs.~(\ref{eq:energyDyOH}) or (\ref{eq:energyErOH})
and solving a linear system for an upper-triangular matrix.
Finally, we note that operator $\hat V$ makes it possible to
construct a Hamiltonian for the rotation of the triatomic molecule near their equilibrium geometry.

We have found it instructive to study the bonds in DyOH and ErOH  by visualizing unit-normalized Kohn-Sham molecular orbitals (MOs) from DFT calculations even though these DFT calculations predict the incorrect order of eigenstate energies. 
Figure \ref{fig:combined}  shows isosurfaces of DyOH and ErOH molecular orbitals with an isovalue of $-0.0025e$\AA$^{-3/2}$ and $+0.0025e$\AA$^{-3/2}$  for the MOs mainly composed of the lanthanide 6s, 4f and 6p atomic orbitals. 
Here, length 1\AA\ is 0.1 nm. The DFT calculations have been performed within the Gaussian-16 package~\cite{g16} and visualized with Gauss-View 6~\cite{q-chem}. The atoms in the molecules are at their equilibrium locations. 
One can immediately observe  that the 6s and 6p MOs easily enclose the Ln, O, H nuclei and that the 4f  orbital has a  radius close to the Ln-O equilibrium separation. In fact, the root-mean-square diameter of the 6s$^2$ molecular orbital is $7a_0$ larger than the equilibrium separation between the lanthanide and hydrogen atom of approximately $3.7 a_0+1.8a_0=5.5a_0$. 

\begin{figure}
\includegraphics[scale=0.28,trim=0 0 0 0,clip]{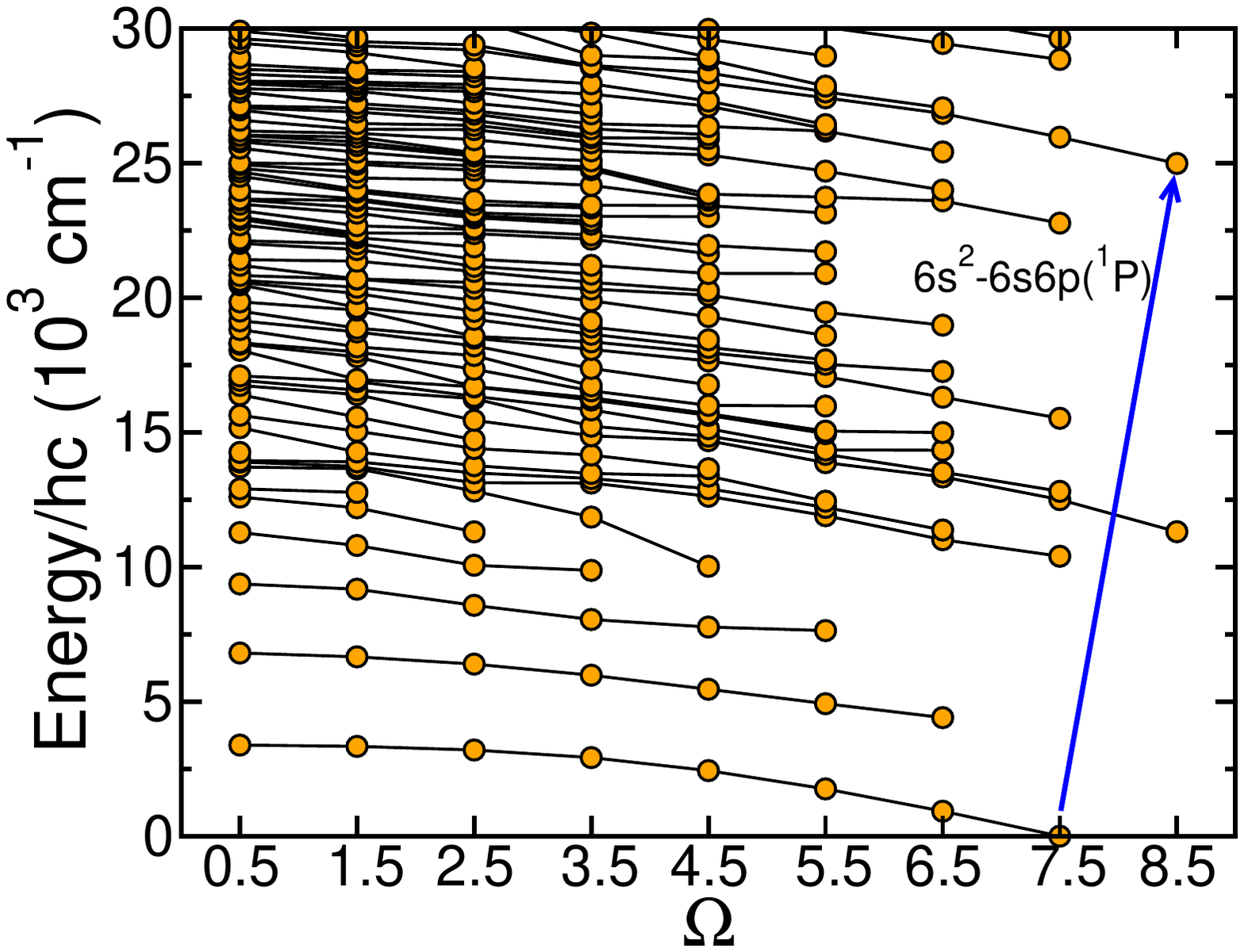}
\caption{ Electronic eigenenergies  (colored circles) of DyOH at their equilibrium, linear geometry as a function of electron projection quantum number $\Omega$. The zero of energy is set at that of the lowest eigenstate.  The excitation energies have been obtained with self-consistent-field  calculations using basis sets that {\it do} include excitations into 6p and 5d molecular orbitals. For the lowest energies, lines connecting the colored circles correspond to states of the same configuration. 
For the higher energies, the lines although sorted by energy, are only guides for the eye. The blue arrow highlights the transition from $\Omega=15/2$ to $\Omega'=17/2$ between the $\rm 6s^2$ and $\rm6s6p$ states. It has a electric dipole moment of $1.79 ea_0$.}
\label{fig:energy}
\end{figure}

The excited level structure of the DyOH and ErOH molecules obtained using electronic structure calculations with a limited set of basis functions has been shown in Figs.~\ref{fig:LnOH}(a) and (b). 
Figure \ref{fig:energy} shows the electronic eigenenergies of DyOH at its equilibrium geometry up to $hc\times 30\,000$~cm$^{-1}$ above its ground state as functions of $\Omega$
when we do include the 5d and 6p molecular orbitals.   Below $hc\times 10\,000$~cm$^{-1}$, states  can still be assigned
as belonging to the $4{\rm f}^{9} 6s{^2}+2p^6$ configuration. For larger excitation energies and especially for states with a small $\Omega$  mixing among the 6s, 5d, and 6p lanthanide molecular orbitals is strong and state assignment was not possible. For large $\Omega$ we can make assignments and, for example,
the arrow in the figure indicates the $\Omega=15/2$ to $\Omega'=17/2$ transition between the ground $\rm 4f^{9} 6s^2+2p^6$ and excited $\rm4f^{9} 6s6p(^1P)+2p^6$ configurations. 
It has a computed transition electric dipole moment of $1.79 ea_0$.

\subsection{Permanent electric dipole moment}

\begin{figure}
\includegraphics[scale=0.28,trim=0 0 0 0,clip]{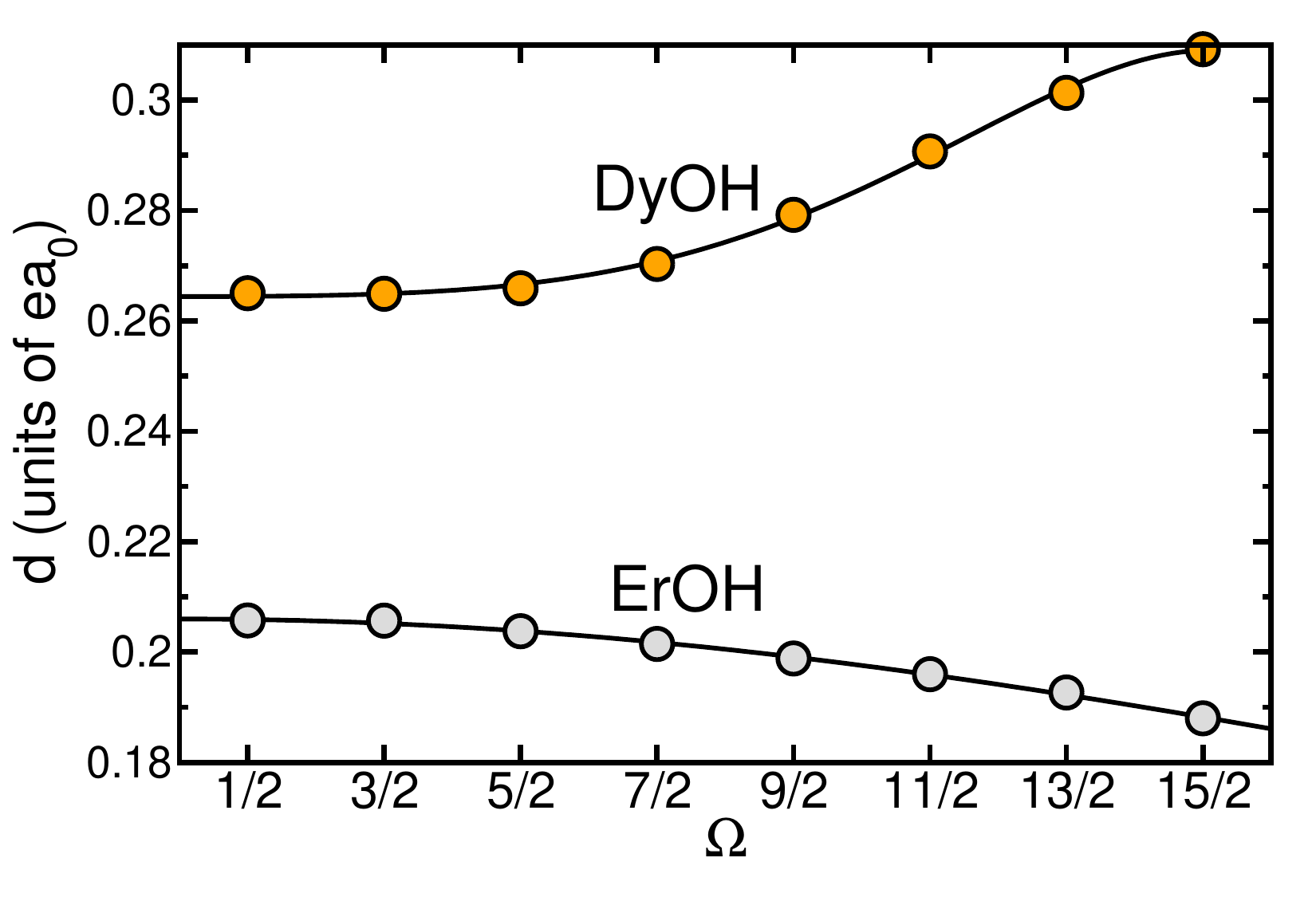}
\caption{Permanent dipole moment in atomic units $ea_0$ as function of projection quantum number $\Omega$
of states of the $4{\rm f}^{n-1}6{\rm s}^2+2{\rm p}^6$ electronic ground-state  configuration of DyOH (orange filled circles) and ErOH (grey filled circles) at their equilibrium linear geometry. Solid black curves are polynomial fits to the data
as described in the text.} 
\label{fig:dipole_moment}
\end{figure}

We have also calculated permanent electronic dipole moments ${\bf d}$ for the  $\Omega$ states of the energetically lowest bundle or group of levels of  configuration ${c=4{\rm f}^{n-1}6{\rm s}^2+2{\rm p}^6}$. Our polar molecules can be controlled with a static electric field ${\bf E}$ via the Hamiltonian $-{\bf d}\cdot {\bf E}$. Figure~\ref{fig:dipole_moment} shows matrix elements $\langle c,+\Omega| d_{z} | c,+\Omega\rangle=\langle c,-\Omega| d_{z} | c,-\Omega\rangle$ for $\Omega>0$ for DyOH and ErOH along the body-fixed $z$ axis directed from the lanthanide atom to the oxygen atom.  These data have been determined for the equilibrium linear geometries and obtained with  the RAS-SCF method using basis sets that include excitations to the 6p and 5d molecular orbitals.

The permanent dipole moments are approximately $0.28ea_0$ and $0.20ea_0$ for DyOH and ErOH, respectively, as shown in Fig.~\ref{fig:dipole_moment}, which indicates that the bond between the
lanthanide atom and hydroxyl is not fully ionic. With $R_{\rm Ln\mhyphen O}\approx3.7a_0$ at the equilibrium geometry of the triatomic molecules, less than 0.1 of an electron charge
has transferred to the hydroxyl.
We believe that this observation does not contradict the finding in the previous section that ground-state electronic configuration   is to good approximation an eigenstate of ${\bf j}^2$
as the closed $6s^2$ molecular orbital envelops the whole molecule as seen in Fig.~\ref{fig:combined}.
For both triatomics, the absolute ground state has the largest permanent dipole moment,
although the $\Omega$ dependence of the dipole moment is weak changing by no more than 20\,\%
for DyOH and significantly less for ErOH.
Finally, we realize that we can write
\begin{eqnarray}
   d_z/(ea_0) &=&  0.264\,395+1.458\,57\times10^{-4}\Omega^2
\\
 && \ \ {}+3.651\,51\times10^{-5}\Omega^4-4.448\,32 \times 10^{-7} \Omega^6 \nonumber
 \end{eqnarray}
for DyOH and 
\begin{eqnarray}
   d_z/(ea_0) &=& 	 0.205\,989
	 -3.531\,68 \times10^{-4}\Omega^2
\end{eqnarray}
for ErOH, where we have omitted the bra and ket notation in the matrix elements for clarity.

Similar to Eq.~(\ref{eq:BOtensor}), we can analyze the $\Omega$ dependence
of the permanent electric dipole moment with spin-spin tensor operators assuming that the total electron
angular momentum is conserved. That is,
\begin{eqnarray}
      -{\bf d}\cdot {\bf E}&=&-d_0 C_1(\hat R)\cdot {\bf E} - d_2 T_1(T_{2}(j,j),C_3(\hat R))\cdot {\bf E}\label{eq:dz} \\
        &&\ - d_4 T_1(T_4(T_{2}(j,j),T_{2}(j,j)),C_5(\hat R))\cdot {\bf E} + \cdots\nonumber
\end{eqnarray}
with coefficients $d_{k}$ and  rank-1 spin-spin operators $T_1(T_{k}(\cdot,\cdot),C_{k+1}(\hat R))$
with even $k=2,4,\cdots$.
In Eq.~(\ref{eq:dz}), the space- or laboratory-fixed electric field ${\bf E}=(E_X,E_Y,E_Z)$ is expressed in ``spherical''
form, that is $E_{\pm 1}=\mp(E_X\pm iE_Y)/\sqrt{2}$ and $E_0=E_Z$,
where  $X$, $Y$, and $Z$ define the space-fixed coordinate axes as before.

The diagonal matrix elements of these spin-spin operators in the basis  $|c,j\Omega\rangle$
and $\hat R=\hat 0$ lead to  polynomials in $\Omega^{2}$ of degree $k/2$.
For example, for $k=2$ we have
\begin{eqnarray}
\lefteqn{ \langle j\Omega| T_{10}(T_{2}(j,j),C_3(\hat R=\hat 0)) |j\Omega\rangle =}\\
&&\langle 10 | 2 3 0 0\rangle \langle j\Omega| T_{20}(j,j) |j\Omega\rangle
=  \frac{3}{\sqrt{35}} \frac{3\Omega^2-j(j+1)}{\sqrt{6}}\,. \nonumber
\end{eqnarray}

\subsection{Magnetic moments and pseudospin Hamiltonians for the ground-state configuration}\label{sec:magmom}

The states of the linear DyOH and ErOH molecules can also be controlled by an external magnetic field ${\bf B}$ through the Zeeman interaction $-\boldsymbol{\mu}\cdot {\bf B}$, where vector operator $\boldsymbol{\mu}$ is the electronic magnetic moment. Within Open-Molcas, we have computed the diagonal and off-diagonal matrix elements of $\mu_\alpha$ of the magnetic moment operator projected along the  body-fixed Cartesian axes $\alpha=x$, $y$, and $z$ for the $2\times8=16$ levels 
of the lowest-energy bundle assigned with configuration ${c=4{\rm f}^{n-1}6{\rm s}^2+2{\rm p}^6}$.
That is, the levels whose energies are shown in Fig.~\ref{fig:LnOH}(c).
 
\begin{figure}
\includegraphics[scale=0.29,trim=0 10 0 0,clip]{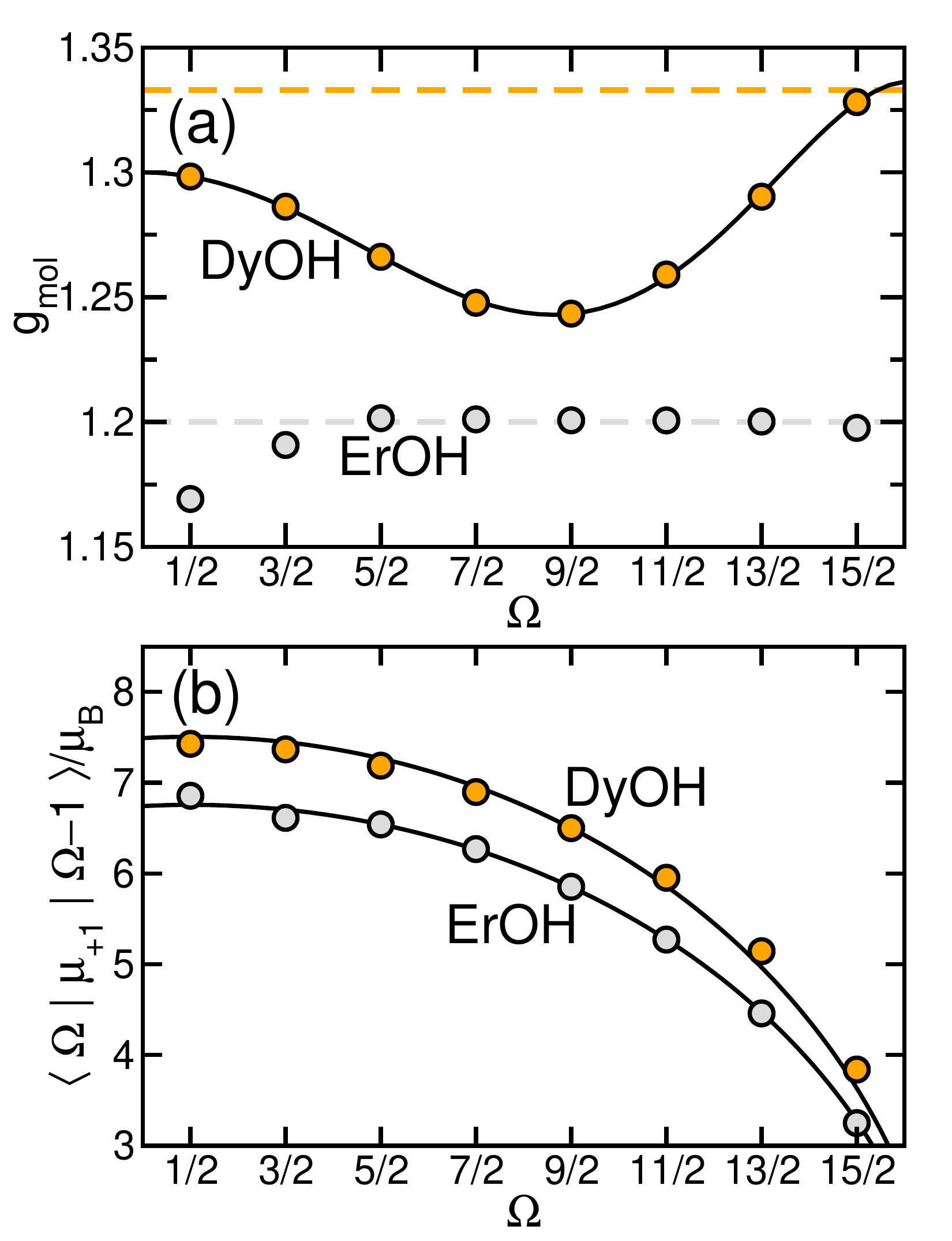}
\caption{Molecular $g$ factors, $g_{\rm mol}$ as defined in the text, (panel a) and transition magnetic  moments (panel b) as functions of $\Omega$ for DyOH (orange filled circles) and ErOH (grey filled circles) at their linear equilibrium geometries along the body-fixed Ln-O axis. $G$-factors of the pseudo-spin are defined in the text.
The DyOH and ErOH molecules are in their  4f$^9$6s$^2$+2p$^6$ and 4f$^{11}$6s$^2$+2p$^6$ ground-state configuration, respectively. 
In panel a)  the dashed orange and grey lines are the experimental $g$ factors of the $j=15/2$ level of the 4f$^9$6s$^2$ and 4f$^{11}$6s$^2$
configurations of  Dy$^+$ and Er$^+$, respectively.
In panel a) the solid black curve is a polynomial in $\Omega^2$ found from a fit to the data  for DyOH,
while the solid black curves in panel b)  correspond to a fit to adjustable parameter $g_1$ times matrix elements of the angular momentum raising operator $j_{+1}/\hbar$.}
\label{fig:gz}
\end{figure}

Following Ref.~\cite{Chibotaru2012} and assuming that the total angular momentum ${\bf j}$ is a  conserved 
operator with quantum number $j=15/2$, it is reasonable to initially propose 
$\boldsymbol{\mu}= g_{\rm mol} \mu_{\rm B} {\bf j}/\hbar$, where dimensionless $g_{\rm mol}$ is a  molecular $g$ factor.
Then because the zero field splittings of the relevant electronic states with different 
$|\Omega|$ are well separated, the diagonal magnetic-moment matrix elements
along the body-fixed $z$ axis, $\langle c,\Omega | \mu_z | c,\Omega\rangle= g_{\rm mol} \mu_{\rm B} \Omega$
will describe the level shifts of the molecular state in a magnetic field absent the rotation of the molecules.

Figure~\ref{fig:gz}(a) shows the computed $g_{\rm mol}$
 as functions of our half-integer $\Omega$ for the ground-state configuration of DyOH and ErOH.
The $g$-factors $g_{\rm mol}$ for both molecules are indeed mostly independent
of $\Omega$ as expected and are even functions of $\Omega$.
The magnetic moment operator  $\mu_{z}$ is thus an odd function of $\Omega$. 
In fact, for DyOH a linear-least-squares fit with the cubic polynomial in $\Omega^2$ to the data in Fig.~\ref{fig:gz}(a) gives
\begin{eqnarray}
	g_{\rm mol}&=& 1.300\,01 -0.006\,794\,86 \,\Omega^2\\
	&&\quad
	+\, 2.35401 \times 10^{-4}\Omega^4 -1.88066\times 10^{-6} \Omega^6\,, \nonumber
\end{eqnarray}
while for ErOH $g_{\rm mol}$ is independent of $\Omega$ for $\Omega>3/2$ but has significant deviations from this behavior for smaller $\Omega$ likely indicating mixing with excited electronic configurations. 
(Polynomial fits to $g_{\rm mol}$ for ErOH do not lead to satisfactory agreements.)

Interestingly,  as also shown in Fig.~\ref{fig:gz}(a) the atomic $g$ factors for the $j=15/2$ level of the 4f$^9$6s$^2$ configuration of 
an electronically excited  Dy$^+$ ion and that of the 4f$^{11}$6s$^2$ configuration of an excited  Er$^+$ ion are 1.333 and 1.200, respectively\cite{Kramida2023}, are in agreement with the molecular $g$ factors. The agreement justifies our Ansatz that
the ground states of DyOH and ErOH have a total molecular electronic angular momentum that is close to $j=15/2$.

As with the relativistic energies in Eq.~(\ref{eq:energyDyOH}), the $\Omega$ dependences can  be described in 
terms of effective spin-spin interactions. The simplest extension for the magnetic moment operator is the rank-1 or vector operator
\begin{equation}
   \frac{ \boldsymbol{\mu}}{\mu_{\rm B}}= g_1  {\bf j}/\hbar+ g_3T_1(T_3(j,T_2(j,j)),C_2(\hat R)) +\cdots
    \label{eq:mu}
\end{equation}
with $g$ factors $g_1$ and $g_3$ in analogy to Eq.~(\ref{eq:BOtensor}). This extension leads to matrix elements for $\mu_z$ 
that are odd  polynomials in $\Omega$.  For example, the spin dependence of the second term on the right hand side of Eq.~(\ref{eq:mu}) equals $\Omega\{ 3j(j+1)-1-5\Omega^2\}$ times a function that solely depends on $j$. For DyOH, more complex rank-1 operators containing spherical harmonics  $C_{4q}(\hat R)$ and $C_{6q}(\hat R)$ can be added to represent
the $\Omega^5$ and $\Omega^7$ dependence of $g_{\rm mol}\Omega$, respectively.

Figure \ref{fig:gz}(b) shows the transition magnetic moments between states with projection quantum numbers $\Omega$ and $\Omega-1$   of the 
$c=4{\rm f}^9$6s$^2$+2p$^6$ configuration of DyOH and the 4f$^{11}$6s$^2$+2p$^6$  configuration of ErOH.
That is, we show matrix elements
\begin{equation}
      \langle c, \Omega | \mu_{+1} | c, {\Omega-1} \rangle/\mu_{\rm B}
\end{equation}
as function of $\Omega\ge 1/2$. 
Assuming that the states have ${j=15/2}$ and $\boldsymbol{\mu}= g_1 \mu_{\rm B} {\bf j}/\hbar$,
we fit these data to  matrix elements of operator $g_1\mu_{\rm B} j_{+1}/\hbar$
with adjustable $g$ factor $g_1$ using that
\begin{eqnarray}
 \langle j \Omega | (j_{+1}/\hbar) | j {\Omega-1} \rangle 
     & =  & \sqrt{\frac{(j-\Omega+1)(j+\Omega)}{2}} 
     \label{eq:mux}
\end{eqnarray}
With $j=15/2$, the fits lead to $g_1=1.32691$ and 1.19430 for DyOH and ErOH, respectively,
close to the values for the $g$-factors of the isolated excited Dy$^+$ and Er$^+$ ions given in two  paragraphs earlier.
Small deviations from the $\Omega$ dependence in Eq.~(\ref{eq:mux}) are due to more complex rank-1 tensors as in Eq.~(\ref{eq:mu}).

\section*{Conclusion}

We have argued that promising applications for ultracold molecules  include performing precision spectroscopy to test the Standard Model of particle physics, creating new types of sensors, advancing quantum information science, simulation of complex exotic materials as well as the promise of quantum control of chemical reactions as each molecule can be prepared in a unique rovibrational quantum state. 

In this paper, we have taken the first steps in understanding the properties of two magnetic Lanthanide monohydroxide
molecules, Dy-OH and Er-OH, in these contexts. We computed ground and excited relativistic electronic energy levels near their linear equilibrium geometry, 
assigned their dominant electronic configuration, and realized that for the ground state an electron from the open submerged 4f shell moves into the 2p$^5$ orbital of OH.
In addition, we showed that the  molecules are both polar and paramagnetic by computing electric and magnetic moments and thus can be simultaneously manipulated by electric and magnetic fields.  Most importantly, we find that
the energetically lowest 16 states form a spin $j=15/2$ system with zero field splittings on the order of a $hc\times 1\,000$~cm$^{-1}$. In fact, this spin system can be understood as being due
to the $j=15/2$ level of the $\rm 4f^9 6s^2$ and $\rm 4f^{11} 6s^2$ configurations of the isolated Dy$^+$ and Er$^+$ ions, respectively. In the future, the understanding gained in this paper  will enable us to determine  rotational and vibrational
of the molecules and hopefully make suggestions for cooling the molecules with  lasers.

\section*{Methods}

We have performed multi-configuration restricted-active-space self-consistent field (RAS-SCF) electronic-structure calculations \cite{Malmqvist1990} with
spin-orbit coupling using by state interaction (RAS-SI) \cite{Malmqvist2002}. 
In the state-interaction approach, the spin-orbit Hamiltonian matrix is built using scalar-relativistic RAS-SCF eigenenergies
of the Douglas-Kroll Hamiltonian and spin-orbit matrix elements among the SCF eigenfunctions.
The resulting low-dimensional $D\times D$ Hamiltonian is diagonalized to obtain relativistic adiabatic potential energies and electronic wavefunctions.
We use the implementation
of the RAS-SCF and RAS-SI methods in OpenMolcas \cite{OpenMolcas} and rely on the built-in ANO-RCC-VQZP electronic basis \cite{Roos2008}.  
For both DyOH and ErOH, we perform  separate non-relativistic RAS-SCF calculations for total electron spin $S=1/2$ (doublets), $3/2$ (quartets), and $5/2$ (sextets).  For  total electron spin $S$, we determine the energetically lowest $D=80$ roots or eigenstates to ensure convergence with respect to couplings due to the spin-orbit interaction. We also use OpenMolcas to determine one-electron properties such as angular-momentum and permanent and transition electric and magnetic dipole moments between relativistic electronic states.

We verify the  RAS-SCF and RAS-SI results with relativistic coupled-cluster calculations for the
electronic ground state and equation-of-motion coupled-cluster calculations for some of the excited states using
the CFOUR package equipped with additional modules \cite{cfour,Matthews2020}. 
That is, we  use relativistic coupled-cluster calculations with single and double excitations (CCSD) \cite{Purvis1982} augmented with non-iterative triples [CCSD(T)] \cite{Raghavachari1989} and equation of motion CCSD \cite{Stanton1993} calculations using the exact-two-component (X2C) Hamiltonian with atomic mean-field integrals (the so-called X2CAMF scheme) \cite{Liu2018,Zhang2022}.  
In the calculations, we use the all-electron uncontracted ANO-RCC-VTZP  basis sets for the lanthanide atoms and the cc-PVTZ-X2C and cc-PVTZ-DK  basis sets for the O and H atoms, respectively. 
The basis set for O is a ``recontracted'' basis for use in the X2CAMF scheme \cite{Dunning1989}.

\vspace*{0.5cm}
{\bf \large{Data availability statement}}\\

The datasets used and/or analyzed during the current study available from the corresponding author on reasonable request.

\vspace*{0.5cm}
{\bf \large{Acknowledgements}}\\

The work at Temple University was funded from the AFOSR, Grant No. FA9550-21-1-0153, the National Science Foundation, Grant No. PHY-2409425, and the Gordon and Betty Moore Foundation. 
The work at the Johns Hopkins University was supported by the National Science Foundation,  Grant No. PHY-2309253.

\bibliography{refs_new_9}

\begin{thebibliography}{45}%
\makeatletter
\providecommand \@ifxundefined [1]{%
 \@ifx{#1\undefined}
}%
\providecommand \@ifnum [1]{%
 \ifnum #1\expandafter \@firstoftwo
 \else \expandafter \@secondoftwo
 \fi
}%
\providecommand \@ifx [1]{%
 \ifx #1\expandafter \@firstoftwo
 \else \expandafter \@secondoftwo
 \fi
}%
\providecommand \natexlab [1]{#1}%
\providecommand \enquote  [1]{``#1''}%
\providecommand \bibnamefont  [1]{#1}%
\providecommand \bibfnamefont [1]{#1}%
\providecommand \citenamefont [1]{#1}%
\providecommand \href@noop [0]{\@secondoftwo}%
\providecommand \href [0]{\begingroup \@sanitize@url \@href}%
\providecommand \@href[1]{\@@startlink{#1}\@@href}%
\providecommand \@@href[1]{\endgroup#1\@@endlink}%
\providecommand \@sanitize@url [0]{\catcode `\\12\catcode `\$12\catcode
  `\&12\catcode `\#12\catcode `\^12\catcode `\_12\catcode `\%12\relax}%
\providecommand \@@startlink[1]{}%
\providecommand \@@endlink[0]{}%
\providecommand \url  [0]{\begingroup\@sanitize@url \@url }%
\providecommand \@url [1]{\endgroup\@href {#1}{\urlprefix }}%
\providecommand \urlprefix  [0]{URL }%
\providecommand \Eprint [0]{\href }%
\providecommand \doibase [0]{http://dx.doi.org/}%
\providecommand \selectlanguage [0]{\@gobble}%
\providecommand \bibinfo  [0]{\@secondoftwo}%
\providecommand \bibfield  [0]{\@secondoftwo}%
\providecommand \translation [1]{[#1]}%
\providecommand \BibitemOpen [0]{}%
\providecommand \bibitemStop [0]{}%
\providecommand \bibitemNoStop [0]{.\EOS\space}%
\providecommand \EOS [0]{\spacefactor3000\relax}%
\providecommand \BibitemShut  [1]{\csname bibitem#1\endcsname}%
\let\auto@bib@innerbib\@empty
\bibitem [{\citenamefont {Lu}\ \emph {et~al.}(2011)\citenamefont {Lu},
  \citenamefont {Burdick}, \citenamefont {Youn},\ and\ \citenamefont
  {Lev}}]{Lu2011}%
  \BibitemOpen
  \bibfield  {author} {\bibinfo {author} {\bibfnamefont {M.}~\bibnamefont
  {Lu}}, \bibinfo {author} {\bibfnamefont {N.~Q.}\ \bibnamefont {Burdick}},
  \bibinfo {author} {\bibfnamefont {S.~H.}\ \bibnamefont {Youn}}, \ and\
  \bibinfo {author} {\bibfnamefont {B.~L.}\ \bibnamefont {Lev}},\ }\bibfield
  {title} {\enquote {\bibinfo {title} {Strongly dipolar {B}ose-{E}instein
  condensate of {D}ysprosium},}\ }\href
  {https://doi.org/10.1103/PhysRevLett.107.190401} {\bibfield  {journal}
  {\bibinfo  {journal} {Phys. Rev. Lett.}\ }\textbf {\bibinfo {volume} {107}},\
  \bibinfo {pages} {190401} (\bibinfo {year} {2011})}\BibitemShut {NoStop}%
\bibitem [{\citenamefont {Lu}\ \emph {et~al.}(2012)\citenamefont {Lu},
  \citenamefont {Burdick},\ and\ \citenamefont {Lev}}]{Lu2012}%
  \BibitemOpen
  \bibfield  {author} {\bibinfo {author} {\bibfnamefont {M.}~\bibnamefont
  {Lu}}, \bibinfo {author} {\bibfnamefont {N.~Q.}\ \bibnamefont {Burdick}}, \
  and\ \bibinfo {author} {\bibfnamefont {B.~L.}\ \bibnamefont {Lev}},\
  }\bibfield  {title} {\enquote {\bibinfo {title} {Quantum degenerate dipolar
  {F}ermi gas},}\ }\href {https://doi.org/10.1103/PhysRevLett.108.215301}
  {\bibfield  {journal} {\bibinfo  {journal} {Phys. Rev. Lett.}\ }\textbf
  {\bibinfo {volume} {108}},\ \bibinfo {pages} {215301} (\bibinfo {year}
  {2012})}\BibitemShut {NoStop}%
\bibitem [{\citenamefont {Frisch}\ \emph {et~al.}(2012)\citenamefont {Frisch},
  \citenamefont {Aikawa}, \citenamefont {Mark}, \citenamefont {Rietzler},
  \citenamefont {Schindler}, \citenamefont {Zupani\ifmmode~\check{c}\else
  \v{c}\fi{}}, \citenamefont {Grimm},\ and\ \citenamefont
  {Ferlaino}}]{Frisch2012}%
  \BibitemOpen
  \bibfield  {author} {\bibinfo {author} {\bibfnamefont {A.}~\bibnamefont
  {Frisch}}, \bibinfo {author} {\bibfnamefont {K.}~\bibnamefont {Aikawa}},
  \bibinfo {author} {\bibfnamefont {M.}~\bibnamefont {Mark}}, \bibinfo {author}
  {\bibfnamefont {A.}~\bibnamefont {Rietzler}}, \bibinfo {author}
  {\bibfnamefont {J.}~\bibnamefont {Schindler}}, \bibinfo {author}
  {\bibfnamefont {E.}~\bibnamefont {Zupani\ifmmode~\check{c}\else \v{c}\fi{}}},
  \bibinfo {author} {\bibfnamefont {R.}~\bibnamefont {Grimm}}, \ and\ \bibinfo
  {author} {\bibfnamefont {F.}~\bibnamefont {Ferlaino}},\ }\bibfield  {title}
  {\enquote {\bibinfo {title} {Narrow-line magneto-optical trap for erbium},}\
  }\href {\doibase 10.1103/PhysRevA.85.051401} {\bibfield  {journal} {\bibinfo
  {journal} {Phys. Rev. A}\ }\textbf {\bibinfo {volume} {85}},\ \bibinfo
  {pages} {051401} (\bibinfo {year} {2012})}\BibitemShut {NoStop}%
\bibitem [{\citenamefont {Aikawa}\ \emph {et~al.}(2012)\citenamefont {Aikawa},
  \citenamefont {Frisch}, \citenamefont {Mark}, \citenamefont {Baier},
  \citenamefont {Rietzler}, \citenamefont {Grimm},\ and\ \citenamefont
  {Ferlaino}}]{Aikawa2012}%
  \BibitemOpen
  \bibfield  {author} {\bibinfo {author} {\bibfnamefont {K.}~\bibnamefont
  {Aikawa}}, \bibinfo {author} {\bibfnamefont {A.}~\bibnamefont {Frisch}},
  \bibinfo {author} {\bibfnamefont {M.}~\bibnamefont {Mark}}, \bibinfo {author}
  {\bibfnamefont {S.}~\bibnamefont {Baier}}, \bibinfo {author} {\bibfnamefont
  {A.}~\bibnamefont {Rietzler}}, \bibinfo {author} {\bibfnamefont
  {R.}~\bibnamefont {Grimm}}, \ and\ \bibinfo {author} {\bibfnamefont
  {F.}~\bibnamefont {Ferlaino}},\ }\bibfield  {title} {\enquote {\bibinfo
  {title} {Bose-einstein condensation of erbium},}\ }\href {\doibase
  10.1103/PhysRevLett.108.210401} {\bibfield  {journal} {\bibinfo  {journal}
  {Phys. Rev. Lett.}\ }\textbf {\bibinfo {volume} {108}},\ \bibinfo {pages}
  {210401} (\bibinfo {year} {2012})}\BibitemShut {NoStop}%
\bibitem [{\citenamefont {Go}\ \emph {et~al.}(2017)\citenamefont {Go},
  \citenamefont {Hanke}, \citenamefont {Buhl}, \citenamefont {Freimuth},
  \citenamefont {Bihlmayer}, \citenamefont {Lee}, \citenamefont {Mokrousov},\
  and\ \citenamefont {Bl\"{u}gel}}]{Go2017}%
  \BibitemOpen
  \bibfield  {author} {\bibinfo {author} {\bibfnamefont {D.}~\bibnamefont
  {Go}}, \bibinfo {author} {\bibfnamefont {J.-P.}\ \bibnamefont {Hanke}},
  \bibinfo {author} {\bibfnamefont {P.~M.}\ \bibnamefont {Buhl}}, \bibinfo
  {author} {\bibfnamefont {F.}~\bibnamefont {Freimuth}}, \bibinfo {author}
  {\bibfnamefont {G.}~\bibnamefont {Bihlmayer}}, \bibinfo {author}
  {\bibfnamefont {H.-W.}\ \bibnamefont {Lee}}, \bibinfo {author} {\bibfnamefont
  {Y.}~\bibnamefont {Mokrousov}}, \ and\ \bibinfo {author} {\bibfnamefont
  {S.}~\bibnamefont {Bl\"{u}gel}},\ }\bibfield  {title} {\enquote {\bibinfo
  {title} {Toward surface orbitronics: {G}iant orbital magnetism from the
  orbital {R}ashba effect at the surface of sp-metals},}\ }\href
  {https://doi.org/10.1038/srep46742} {\bibfield  {journal} {\bibinfo
  {journal} {Scientific Reports}\ }\textbf {\bibinfo {volume} {7}},\ \bibinfo
  {pages} {46742} (\bibinfo {year} {2017})}\BibitemShut {NoStop}%
\bibitem [{\citenamefont {Bernevig}\ \emph {et~al.}(2005)\citenamefont
  {Bernevig}, \citenamefont {Hughes},\ and\ \citenamefont
  {Zhang}}]{Bernevig2005}%
  \BibitemOpen
  \bibfield  {author} {\bibinfo {author} {\bibfnamefont {B.~A.}\ \bibnamefont
  {Bernevig}}, \bibinfo {author} {\bibfnamefont {T.~L.}\ \bibnamefont
  {Hughes}}, \ and\ \bibinfo {author} {\bibfnamefont {S.-C.}\ \bibnamefont
  {Zhang}},\ }\bibfield  {title} {\enquote {\bibinfo {title} {Orbitronics: The
  intrinsic orbital current in $p$-doped silicon},}\ }\href {\doibase
  10.1103/PhysRevLett.95.066601} {\bibfield  {journal} {\bibinfo  {journal}
  {Phys. Rev. Lett.}\ }\textbf {\bibinfo {volume} {95}},\ \bibinfo {pages}
  {066601} (\bibinfo {year} {2005})}\BibitemShut {NoStop}%
\bibitem [{\citenamefont {Isaev}\ \emph {et~al.}(2017)\citenamefont {Isaev},
  \citenamefont {Zaitsevskii},\ and\ \citenamefont {Eliav}}]{Isaev2017}%
  \BibitemOpen
  \bibfield  {author} {\bibinfo {author} {\bibfnamefont {T.~A.}\ \bibnamefont
  {Isaev}}, \bibinfo {author} {\bibfnamefont {A.~V.}\ \bibnamefont
  {Zaitsevskii}}, \ and\ \bibinfo {author} {\bibfnamefont {E.}~\bibnamefont
  {Eliav}},\ }\bibfield  {title} {\enquote {\bibinfo {title} {Laser-coolable
  polyatomic molecules with heavy nuclei},}\ }\href {\doibase
  10.1088/1361-6455/aa8f34} {\bibfield  {journal} {\bibinfo  {journal} {J.
  Phys. B}\ }\textbf {\bibinfo {volume} {50}},\ \bibinfo {pages} {225101}
  (\bibinfo {year} {2017})}\BibitemShut {NoStop}%
\bibitem [{\citenamefont {Kozyryev}\ \emph {et~al.}(2017)\citenamefont
  {Kozyryev}, \citenamefont {Baum}, \citenamefont {Matsuda}, \citenamefont
  {Augenbraun}, \citenamefont {Anderegg}, \citenamefont {Sedlack},\ and\
  \citenamefont {Doyle}}]{Kozyryev2017}%
  \BibitemOpen
  \bibfield  {author} {\bibinfo {author} {\bibfnamefont {I.}~\bibnamefont
  {Kozyryev}}, \bibinfo {author} {\bibfnamefont {L.}~\bibnamefont {Baum}},
  \bibinfo {author} {\bibfnamefont {K.}~\bibnamefont {Matsuda}}, \bibinfo
  {author} {\bibfnamefont {B.~L.}\ \bibnamefont {Augenbraun}}, \bibinfo
  {author} {\bibfnamefont {L.}~\bibnamefont {Anderegg}}, \bibinfo {author}
  {\bibfnamefont {A.~P.}\ \bibnamefont {Sedlack}}, \ and\ \bibinfo {author}
  {\bibfnamefont {J.~M.}\ \bibnamefont {Doyle}},\ }\bibfield  {title} {\enquote
  {\bibinfo {title} {Sisyphus laser cooling of a polyatomic molecule},}\ }\href
  {\doibase 10.1103/PhysRevLett.118.173201} {\bibfield  {journal} {\bibinfo
  {journal} {Phys. Rev. Lett.}\ }\textbf {\bibinfo {volume} {118}},\ \bibinfo
  {pages} {173201} (\bibinfo {year} {2017})}\BibitemShut {NoStop}%
\bibitem [{\citenamefont {O'Rourke}\ and\ \citenamefont
  {Hutzler}(2019)}]{Hutzler2019}%
  \BibitemOpen
  \bibfield  {author} {\bibinfo {author} {\bibfnamefont {M.~J.}\ \bibnamefont
  {O'Rourke}}\ and\ \bibinfo {author} {\bibfnamefont {N.~R.}\ \bibnamefont
  {Hutzler}},\ }\bibfield  {title} {\enquote {\bibinfo {title} {Hypermetallic
  polar molecules for precision measurements},}\ }\href {\doibase
  10.1103/PhysRevA.100.022502} {\bibfield  {journal} {\bibinfo  {journal}
  {Phys. Rev. A}\ }\textbf {\bibinfo {volume} {100}},\ \bibinfo {pages}
  {022502} (\bibinfo {year} {2019})}\BibitemShut {NoStop}%
\bibitem [{\citenamefont {Hutzler}(2020)}]{Hutzler2020}%
  \BibitemOpen
  \bibfield  {author} {\bibinfo {author} {\bibfnamefont {N.~R.}\ \bibnamefont
  {Hutzler}},\ }\bibfield  {title} {\enquote {\bibinfo {title} {Polyatomic
  molecules as quantum sensors for fundamental physics},}\ }\href {\doibase
  10.1088/2058-9565/abb9c5} {\bibfield  {journal} {\bibinfo  {journal} {Quantum
  Sci. Technol.}\ }\textbf {\bibinfo {volume} {5}},\ \bibinfo {pages} {044011}
  (\bibinfo {year} {2020})}\BibitemShut {NoStop}%
\bibitem [{\citenamefont {Augenbraun}\ \emph
  {et~al.}(2020{\natexlab{a}})\citenamefont {Augenbraun}, \citenamefont
  {Doyle}, \citenamefont {Zelevinsky},\ and\ \citenamefont
  {Kozyryev}}]{BAugenbraun2020}%
  \BibitemOpen
  \bibfield  {author} {\bibinfo {author} {\bibfnamefont {B.~L.}\ \bibnamefont
  {Augenbraun}}, \bibinfo {author} {\bibfnamefont {J.~M.}\ \bibnamefont
  {Doyle}}, \bibinfo {author} {\bibfnamefont {T.}~\bibnamefont {Zelevinsky}}, \
  and\ \bibinfo {author} {\bibfnamefont {I.}~\bibnamefont {Kozyryev}},\
  }\bibfield  {title} {\enquote {\bibinfo {title} {Molecular asymmetry and
  optical cycling: Laser cooling asymmetric top molecules},}\ }\href {\doibase
  10.1103/PhysRevX.10.031022} {\bibfield  {journal} {\bibinfo  {journal} {Phys.
  Rev. X}\ }\textbf {\bibinfo {volume} {10}},\ \bibinfo {pages} {031022}
  (\bibinfo {year} {2020}{\natexlab{a}})}\BibitemShut {NoStop}%
\bibitem [{\citenamefont {Chen}\ \emph {et~al.}(2024)\citenamefont {Chen},
  \citenamefont {Zhang}, \citenamefont {Cheng}, \citenamefont {Ng},
  \citenamefont {Malbrunot-Ettenauer}, \citenamefont {Flambaum}, \citenamefont
  {Lasner}, \citenamefont {Doyle}, \citenamefont {Yu}, \citenamefont {Conn},
  \citenamefont {Zhang}, \citenamefont {Hutzler}, \citenamefont {Jayich},
  \citenamefont {Augenbraun},\ and\ \citenamefont {DeMille}}]{Chen2024}%
  \BibitemOpen
  \bibfield  {author} {\bibinfo {author} {\bibfnamefont {T.}~\bibnamefont
  {Chen}}, \bibinfo {author} {\bibfnamefont {C.}~\bibnamefont {Zhang}},
  \bibinfo {author} {\bibfnamefont {L.}~\bibnamefont {Cheng}}, \bibinfo
  {author} {\bibfnamefont {K.~B.}\ \bibnamefont {Ng}}, \bibinfo {author}
  {\bibfnamefont {S.}~\bibnamefont {Malbrunot-Ettenauer}}, \bibinfo {author}
  {\bibfnamefont {V.~V.}\ \bibnamefont {Flambaum}}, \bibinfo {author}
  {\bibfnamefont {Z.}~\bibnamefont {Lasner}}, \bibinfo {author} {\bibfnamefont
  {J.~M.}\ \bibnamefont {Doyle}}, \bibinfo {author} {\bibfnamefont
  {P.}~\bibnamefont {Yu}}, \bibinfo {author} {\bibfnamefont {C.~J.}\
  \bibnamefont {Conn}}, \bibinfo {author} {\bibfnamefont {C.}~\bibnamefont
  {Zhang}}, \bibinfo {author} {\bibfnamefont {N.~R.}\ \bibnamefont {Hutzler}},
  \bibinfo {author} {\bibfnamefont {A.~M.}\ \bibnamefont {Jayich}}, \bibinfo
  {author} {\bibfnamefont {B.}~\bibnamefont {Augenbraun}}, \ and\ \bibinfo
  {author} {\bibfnamefont {D.}~\bibnamefont {DeMille}},\ }\bibfield  {title}
  {\enquote {\bibinfo {title} {Relativistic exact two-component coupled-cluster
  study of molecular sensitivity factors for nuclear {S}chiff moments},}\
  }\href {\doibase 10.1021/acs.jpca.4c02640} {\bibfield  {journal} {\bibinfo
  {journal} {J. Phys. Chem. A}\ }\textbf {\bibinfo {volume} {128}},\ \bibinfo
  {pages} {6540--6554} (\bibinfo {year} {2024})}\BibitemShut {NoStop}%
\bibitem [{\citenamefont {Auerbach}\ \emph {et~al.}(1996)\citenamefont
  {Auerbach}, \citenamefont {Flambaum},\ and\ \citenamefont
  {Spevak}}]{Auerbach1996}%
  \BibitemOpen
  \bibfield  {author} {\bibinfo {author} {\bibfnamefont {N.}~\bibnamefont
  {Auerbach}}, \bibinfo {author} {\bibfnamefont {V.~V.}\ \bibnamefont
  {Flambaum}}, \ and\ \bibinfo {author} {\bibfnamefont {V.}~\bibnamefont
  {Spevak}},\ }\bibfield  {title} {\enquote {\bibinfo {title} {Collective {T}-
  and {P}-odd electromagnetic moments in nuclei with octupole deformations},}\
  }\href {\doibase 10.1103/PhysRevLett.76.4316} {\bibfield  {journal} {\bibinfo
   {journal} {Phys. Rev. Lett.}\ }\textbf {\bibinfo {volume} {76}},\ \bibinfo
  {pages} {4316--4319} (\bibinfo {year} {1996})}\BibitemShut {NoStop}%
\bibitem [{\citenamefont {Dobaczewski}\ and\ \citenamefont
  {Engel}(2005)}]{Dobaczewski2005}%
  \BibitemOpen
  \bibfield  {author} {\bibinfo {author} {\bibfnamefont {J.}~\bibnamefont
  {Dobaczewski}}\ and\ \bibinfo {author} {\bibfnamefont {J.}~\bibnamefont
  {Engel}},\ }\bibfield  {title} {\enquote {\bibinfo {title} {Nuclear
  time-reversal violation and the {S}chiff moment of $^{225}\mathrm{Ra}$},}\
  }\href {\doibase 10.1103/PhysRevLett.94.232502} {\bibfield  {journal}
  {\bibinfo  {journal} {Phys. Rev. Lett.}\ }\textbf {\bibinfo {volume} {94}},\
  \bibinfo {pages} {232502} (\bibinfo {year} {2005})}\BibitemShut {NoStop}%
\bibitem [{\citenamefont {Skripnikov}\ \emph {et~al.}(2020)\citenamefont
  {Skripnikov}, \citenamefont {Mosyagin}, \citenamefont {Titov},\ and\
  \citenamefont {Flambaum}}]{Skripnikov2020}%
  \BibitemOpen
  \bibfield  {author} {\bibinfo {author} {\bibfnamefont {L.~V.}\ \bibnamefont
  {Skripnikov}}, \bibinfo {author} {\bibfnamefont {N.~S.}\ \bibnamefont
  {Mosyagin}}, \bibinfo {author} {\bibfnamefont {A.~V.}\ \bibnamefont {Titov}},
  \ and\ \bibinfo {author} {\bibfnamefont {V.~V.}\ \bibnamefont {Flambaum}},\
  }\bibfield  {title} {\enquote {\bibinfo {title} {Actinide and lanthanide
  molecules to search for strong {CP}-violation},}\ }\href {\doibase
  10.1039/D0CP01989E} {\bibfield  {journal} {\bibinfo  {journal} {Phys. Chem.
  Chem. Phys.}\ }\textbf {\bibinfo {volume} {22}},\ \bibinfo {pages}
  {18374--18380} (\bibinfo {year} {2020})}\BibitemShut {NoStop}%
\bibitem [{\citenamefont {Yu}\ and\ \citenamefont {Hutzler}(2021)}]{Yu2021}%
  \BibitemOpen
  \bibfield  {author} {\bibinfo {author} {\bibfnamefont {P.}~\bibnamefont
  {Yu}}\ and\ \bibinfo {author} {\bibfnamefont {N.~R.}\ \bibnamefont
  {Hutzler}},\ }\bibfield  {title} {\enquote {\bibinfo {title} {Probing
  fundamental symmetries of deformed nuclei in symmetric top molecules},}\
  }\href {\doibase 10.1103/PhysRevLett.126.023003} {\bibfield  {journal}
  {\bibinfo  {journal} {Phys. Rev. Lett.}\ }\textbf {\bibinfo {volume} {126}},\
  \bibinfo {pages} {023003} (\bibinfo {year} {2021})}\BibitemShut {NoStop}%
\bibitem [{\citenamefont {Flambaum}\ and\ \citenamefont
  {Mansour}(2022)}]{Flambaum2022}%
  \BibitemOpen
  \bibfield  {author} {\bibinfo {author} {\bibfnamefont {V.~V.}\ \bibnamefont
  {Flambaum}}\ and\ \bibinfo {author} {\bibfnamefont {A.~J.}\ \bibnamefont
  {Mansour}},\ }\bibfield  {title} {\enquote {\bibinfo {title} {Enhanced
  magnetic quadrupole moments in nuclei with octupole deformation and their
  {CP}-violating effects in molecules},}\ }\href {\doibase
  10.1103/PhysRevC.105.065503} {\bibfield  {journal} {\bibinfo  {journal}
  {Phys. Rev. C}\ }\textbf {\bibinfo {volume} {105}},\ \bibinfo {pages}
  {065503} (\bibinfo {year} {2022})}\BibitemShut {NoStop}%
\bibitem [{\citenamefont {Sishkov}\ \emph {et~al.}(1984)\citenamefont
  {Sishkov}, \citenamefont {Flambaum},\ and\ \citenamefont
  {Khriplovich}}]{Sushkov1984}%
  \BibitemOpen
  \bibfield  {author} {\bibinfo {author} {\bibfnamefont {O.~P.}\ \bibnamefont
  {Sishkov}}, \bibinfo {author} {\bibfnamefont {V.~V.}\ \bibnamefont
  {Flambaum}}, \ and\ \bibinfo {author} {\bibfnamefont {I.~B.}\ \bibnamefont
  {Khriplovich}},\ }\bibfield  {title} {\enquote {\bibinfo {title} {Possibility
  of investigation {P}- and {T}-odd nuclear forces in atomic and molecular
  experiments},}\ }\href {https://api.semanticscholar.org/CorpusID:115533030}
  {\bibfield  {journal} {\bibinfo  {journal} {Sov. Phys. JETP}\ }\textbf
  {\bibinfo {volume} {60}},\ \bibinfo {pages} {873--883} (\bibinfo {year}
  {1984})}\BibitemShut {NoStop}%
\bibitem [{\citenamefont {Flambaum}(1994)}]{Flambaum1994}%
  \BibitemOpen
  \bibfield  {author} {\bibinfo {author} {\bibfnamefont {V.~V.}\ \bibnamefont
  {Flambaum}},\ }\bibfield  {title} {\enquote {\bibinfo {title} {Spin hedgehog
  and collective magnetic quadrupole moments induced by parity and time
  invariance violating interaction},}\ }\href {\doibase
  https://doi.org/10.1016/0370-2693(94)90646-7} {\bibfield  {journal} {\bibinfo
   {journal} {Phys. Lett. B}\ }\textbf {\bibinfo {volume} {320}},\ \bibinfo
  {pages} {211--215} (\bibinfo {year} {1994})}\BibitemShut {NoStop}%
\bibitem [{\citenamefont {Flambaum}\ \emph {et~al.}(2014)\citenamefont
  {Flambaum}, \citenamefont {DeMille},\ and\ \citenamefont
  {Kozlov}}]{Flambaum2014}%
  \BibitemOpen
  \bibfield  {author} {\bibinfo {author} {\bibfnamefont {V.~V.}\ \bibnamefont
  {Flambaum}}, \bibinfo {author} {\bibfnamefont {D.}~\bibnamefont {DeMille}}, \
  and\ \bibinfo {author} {\bibfnamefont {M.~G.}\ \bibnamefont {Kozlov}},\
  }\bibfield  {title} {\enquote {\bibinfo {title} {Time-reversal symmetry
  violation in molecules induced by nuclear magnetic quadrupole moments},}\
  }\href {\doibase 10.1103/PhysRevLett.113.103003} {\bibfield  {journal}
  {\bibinfo  {journal} {Phys. Rev. Lett.}\ }\textbf {\bibinfo {volume} {113}},\
  \bibinfo {pages} {103003} (\bibinfo {year} {2014})}\BibitemShut {NoStop}%
\bibitem [{\citenamefont {McClelland}\ and\ \citenamefont
  {Hanssen}(2006)}]{McClelland2006}%
  \BibitemOpen
  \bibfield  {author} {\bibinfo {author} {\bibfnamefont {J.~J.}\ \bibnamefont
  {McClelland}}\ and\ \bibinfo {author} {\bibfnamefont {J.~L.}\ \bibnamefont
  {Hanssen}},\ }\bibfield  {title} {\enquote {\bibinfo {title} {Laser cooling
  without repumping: A magneto-optical trap for erbium atoms},}\ }\href
  {\doibase 10.1103/PhysRevLett.96.143005} {\bibfield  {journal} {\bibinfo
  {journal} {Phys. Rev. Lett.}\ }\textbf {\bibinfo {volume} {96}},\ \bibinfo
  {pages} {143005} (\bibinfo {year} {2006})}\BibitemShut {NoStop}%
\bibitem [{\citenamefont {Lu}\ \emph {et~al.}(2010)\citenamefont {Lu},
  \citenamefont {Youn},\ and\ \citenamefont {Lev}}]{Lu2010}%
  \BibitemOpen
  \bibfield  {author} {\bibinfo {author} {\bibfnamefont {M.}~\bibnamefont
  {Lu}}, \bibinfo {author} {\bibfnamefont {S.~H.}\ \bibnamefont {Youn}}, \ and\
  \bibinfo {author} {\bibfnamefont {B.~L.}\ \bibnamefont {Lev}},\ }\bibfield
  {title} {\enquote {\bibinfo {title} {Trapping ultracold dysprosium: A highly
  magnetic gas for dipolar physics},}\ }\href {\doibase
  10.1103/PhysRevLett.104.063001} {\bibfield  {journal} {\bibinfo  {journal}
  {Phys. Rev. Lett.}\ }\textbf {\bibinfo {volume} {104}},\ \bibinfo {pages}
  {063001} (\bibinfo {year} {2010})}\BibitemShut {NoStop}%
\bibitem [{\citenamefont {Vilas}\ \emph {et~al.}(2024)\citenamefont {Vilas},
  \citenamefont {Robichaud}, \citenamefont {Hallas}, \citenamefont {Li},
  \citenamefont {Anderegg},\ and\ \citenamefont {Doyle}}]{Vilas2024}%
  \BibitemOpen
  \bibfield  {author} {\bibinfo {author} {\bibfnamefont {N.~B.}\ \bibnamefont
  {Vilas}}, \bibinfo {author} {\bibfnamefont {P.}~\bibnamefont {Robichaud}},
  \bibinfo {author} {\bibfnamefont {C.}~\bibnamefont {Hallas}}, \bibinfo
  {author} {\bibfnamefont {G.~K.}\ \bibnamefont {Li}}, \bibinfo {author}
  {\bibfnamefont {L.}~\bibnamefont {Anderegg}}, \ and\ \bibinfo {author}
  {\bibfnamefont {J.~M.}\ \bibnamefont {Doyle}},\ }\bibfield  {title} {\enquote
  {\bibinfo {title} {An optical tweezer array of ultracold polyatomic
  molecules},}\ }\href {\doibase 10.1038/s41586-024-07199-1} {\bibfield
  {journal} {\bibinfo  {journal} {Nature}\ }\textbf {\bibinfo {volume} {628}},\
  \bibinfo {pages} {282--286} (\bibinfo {year} {2024})}\BibitemShut {NoStop}%
\bibitem [{\citenamefont {Augenbraun}\ \emph
  {et~al.}(2020{\natexlab{b}})\citenamefont {Augenbraun}, \citenamefont
  {Lasner}, \citenamefont {Frenett}, \citenamefont {Sawaoka}, \citenamefont
  {Miller}, \citenamefont {Steimle},\ and\ \citenamefont
  {Doyle}}]{BAugenbraun2020a}%
  \BibitemOpen
  \bibfield  {author} {\bibinfo {author} {\bibfnamefont {B.~L.}\ \bibnamefont
  {Augenbraun}}, \bibinfo {author} {\bibfnamefont {Z.~D.}\ \bibnamefont
  {Lasner}}, \bibinfo {author} {\bibfnamefont {A.}~\bibnamefont {Frenett}},
  \bibinfo {author} {\bibfnamefont {H.}~\bibnamefont {Sawaoka}}, \bibinfo
  {author} {\bibfnamefont {C.}~\bibnamefont {Miller}}, \bibinfo {author}
  {\bibfnamefont {T.~C.}\ \bibnamefont {Steimle}}, \ and\ \bibinfo {author}
  {\bibfnamefont {J.~M.}\ \bibnamefont {Doyle}},\ }\bibfield  {title} {\enquote
  {\bibinfo {title} {Laser-cooled polyatomic molecules for improved electron
  electric dipole moment searches},}\ }\href {\doibase
  10.1088/1367-2630/ab687b} {\bibfield  {journal} {\bibinfo  {journal} {New J.
  Phys.}\ }\textbf {\bibinfo {volume} {22}},\ \bibinfo {pages} {022003}
  (\bibinfo {year} {2020}{\natexlab{b}})}\BibitemShut {NoStop}%
\bibitem [{\citenamefont {Harb}\ \emph {et~al.}(2019)\citenamefont {Harb},
  \citenamefont {Thompson},\ and\ \citenamefont {Hratchian}}]{Harb2019}%
  \BibitemOpen
  \bibfield  {author} {\bibinfo {author} {\bibfnamefont {H.}~\bibnamefont
  {Harb}}, \bibinfo {author} {\bibfnamefont {L.~M.}\ \bibnamefont {Thompson}},
  \ and\ \bibinfo {author} {\bibfnamefont {H.~P.}\ \bibnamefont {Hratchian}},\
  }\bibfield  {title} {\enquote {\bibinfo {title} {On the linear geometry of
  lanthanide hydroxide ({L}n-{OH}, {L}n = {L}a-{L}u)},}\ }\href {\doibase
  10.1039/C9CP01560D} {\bibfield  {journal} {\bibinfo  {journal} {Phys. Chem.
  Chem. Phys.}\ }\textbf {\bibinfo {volume} {21}},\ \bibinfo {pages}
  {21890--21897} (\bibinfo {year} {2019})}\BibitemShut {NoStop}%
\bibitem [{\citenamefont {Aquilante}\ \emph {et~al.}(2020)\citenamefont
  {Aquilante}, \citenamefont {Autschbach}, \citenamefont {Baiardi},
  \citenamefont {Battaglia}, \citenamefont {Borin}, \citenamefont {Chibotaru},
  \citenamefont {Conti}, \citenamefont {De~Vico}, \citenamefont {Delcey},
  \citenamefont {Fdez.~Galv\'an}, \citenamefont {Ferr\'e}, \citenamefont
  {Freitag}, \citenamefont {Garavelli}, \citenamefont {Gong}, \citenamefont
  {Knecht}, \citenamefont {Larsson}, \citenamefont {Lindh}, \citenamefont
  {Lundberg}, \citenamefont {Malmqvist}, \citenamefont {Nenov}, \citenamefont
  {Norell}, \citenamefont {Odelius}, \citenamefont {Olivucci}, \citenamefont
  {Pedersen}, \citenamefont {Pedraza-Gonz\'alez}, \citenamefont {Phung},
  \citenamefont {Pierloot}, \citenamefont {Reiher}, \citenamefont {Schapiro},
  \citenamefont {Segarra-Marti}, \citenamefont {Segatta}, \citenamefont
  {Seijo}, \citenamefont {Sen}, \citenamefont {Sergentu}, \citenamefont
  {Stein}, \citenamefont {Ungur}, \citenamefont {Vacher}, \citenamefont
  {Valentini},\ and\ \citenamefont {Veryazov}}]{OpenMolcas}%
  \BibitemOpen
  \bibfield  {author} {\bibinfo {author} {\bibfnamefont {F.}~\bibnamefont
  {Aquilante}}, \bibinfo {author} {\bibfnamefont {J.}~\bibnamefont
  {Autschbach}}, \bibinfo {author} {\bibfnamefont {A.}~\bibnamefont {Baiardi}},
  \bibinfo {author} {\bibfnamefont {S.}~\bibnamefont {Battaglia}}, \bibinfo
  {author} {\bibfnamefont {V.~A.}\ \bibnamefont {Borin}}, \bibinfo {author}
  {\bibfnamefont {L.~F.}\ \bibnamefont {Chibotaru}}, \bibinfo {author}
  {\bibfnamefont {I.}~\bibnamefont {Conti}}, \bibinfo {author} {\bibfnamefont
  {L.}~\bibnamefont {De~Vico}}, \bibinfo {author} {\bibfnamefont
  {M.}~\bibnamefont {Delcey}}, \bibinfo {author} {\bibfnamefont
  {I.}~\bibnamefont {Fdez.~Galv\'an}}, \bibinfo {author} {\bibfnamefont
  {N.}~\bibnamefont {Ferr\'e}}, \bibinfo {author} {\bibfnamefont
  {L.}~\bibnamefont {Freitag}}, \bibinfo {author} {\bibfnamefont
  {M.}~\bibnamefont {Garavelli}}, \bibinfo {author} {\bibfnamefont
  {X.}~\bibnamefont {Gong}}, \bibinfo {author} {\bibfnamefont {S.}~\bibnamefont
  {Knecht}}, \bibinfo {author} {\bibfnamefont {E.~D.}\ \bibnamefont {Larsson}},
  \bibinfo {author} {\bibfnamefont {R.}~\bibnamefont {Lindh}}, \bibinfo
  {author} {\bibfnamefont {M.}~\bibnamefont {Lundberg}}, \bibinfo {author}
  {\bibfnamefont {P.-\AA.}\ \bibnamefont {Malmqvist}}, \bibinfo {author}
  {\bibfnamefont {A.}~\bibnamefont {Nenov}}, \bibinfo {author} {\bibfnamefont
  {J.}~\bibnamefont {Norell}}, \bibinfo {author} {\bibfnamefont
  {M.}~\bibnamefont {Odelius}}, \bibinfo {author} {\bibfnamefont
  {M.}~\bibnamefont {Olivucci}}, \bibinfo {author} {\bibfnamefont {T.~B.}\
  \bibnamefont {Pedersen}}, \bibinfo {author} {\bibfnamefont {L.}~\bibnamefont
  {Pedraza-Gonz\'alez}}, \bibinfo {author} {\bibfnamefont {Q.~M.}\ \bibnamefont
  {Phung}}, \bibinfo {author} {\bibfnamefont {K.}~\bibnamefont {Pierloot}},
  \bibinfo {author} {\bibfnamefont {M.}~\bibnamefont {Reiher}}, \bibinfo
  {author} {\bibfnamefont {I.}~\bibnamefont {Schapiro}}, \bibinfo {author}
  {\bibfnamefont {J.}~\bibnamefont {Segarra-Marti}}, \bibinfo {author}
  {\bibfnamefont {F.}~\bibnamefont {Segatta}}, \bibinfo {author} {\bibfnamefont
  {L.}~\bibnamefont {Seijo}}, \bibinfo {author} {\bibfnamefont
  {S.}~\bibnamefont {Sen}}, \bibinfo {author} {\bibfnamefont {D.~C.}\
  \bibnamefont {Sergentu}}, \bibinfo {author} {\bibfnamefont {C.~J.}\
  \bibnamefont {Stein}}, \bibinfo {author} {\bibfnamefont {L.}~\bibnamefont
  {Ungur}}, \bibinfo {author} {\bibfnamefont {M.}~\bibnamefont {Vacher}},
  \bibinfo {author} {\bibfnamefont {A.}~\bibnamefont {Valentini}}, \ and\
  \bibinfo {author} {\bibfnamefont {V.}~\bibnamefont {Veryazov}},\ }\bibfield
  {title} {\enquote {\bibinfo {title} {Modern quantum chemistry with
  [{O}pen]{M}olcas},}\ }\href {\doibase 10.1063/5.0004835} {\bibfield
  {journal} {\bibinfo  {journal} {J. Chem. Phys.}\ }\textbf {\bibinfo {volume}
  {152}},\ \bibinfo {pages} {214117--214125} (\bibinfo {year}
  {2020})}\BibitemShut {NoStop}%
\bibitem [{\citenamefont {Stanton}\ \emph {et~al.}()\citenamefont {Stanton},
  \citenamefont {Gauss}, \citenamefont {Cheng}, \citenamefont {Harding},
  \citenamefont {Matthews},\ and\ \citenamefont {Szalay}}]{cfour}%
  \BibitemOpen
  \bibfield  {author} {\bibinfo {author} {\bibfnamefont {J.~F.}\ \bibnamefont
  {Stanton}}, \bibinfo {author} {\bibfnamefont {J.}~\bibnamefont {Gauss}},
  \bibinfo {author} {\bibfnamefont {L.}~\bibnamefont {Cheng}}, \bibinfo
  {author} {\bibfnamefont {M.~E.}\ \bibnamefont {Harding}}, \bibinfo {author}
  {\bibfnamefont {D.~A.}\ \bibnamefont {Matthews}}, \ and\ \bibinfo {author}
  {\bibfnamefont {P.~G.}\ \bibnamefont {Szalay}},\ }\href@noop {} {}\bibinfo
  {note} {{W}ith contributions from {A}. {A}sthana, {A}.{A}. {A}uer, {R}.{J}.
  {B}artlett, {U}. {B}enedikt, {C}. {B}erger, {D}.{E}. {B}ernholdt, {S}.
  {B}laschke, {Y}. {J}. {B}omble, {S}. {B}urger, {O}. {C}hristiansen, {D}.
  {D}atta, {F}. {E}ngel, {R}. {F}aber, {J}. {G}reiner, {M}. {H}eckert, {O}.
  {H}eun, {M}. Hilgenberg, {C}. {H}uber, {T}.-{C}. {J}agau, {D}. {J}onsson,
  {J}. {J}us{\'e}lius, {T}. Kirsch, {M}.-{P}. {K}itsaras, {K}. {K}lein,
  {G}.{M}. {K}opper, {W}.{J}. {L}auderdale, {F}. {L}ipparini, {J}. {L}iu, {T}.
  {M}etzroth, {L}.{A}. {M}{\"u}ck, {D}.{P}. {O}'{N}eill, {T}. {N}ottoli, {J}.
  {O}swald, {D}.{R}. {P}rice, {E}. {P}rochnow, {C}. {P}uzzarini, {K}. {R}uud,
  {F}. {S}chiffmann, {W}. {S}chwalbach, {C}. {S}immons, {S}. {S}topkowicz, {A}.
  {T}ajti, {T.} Uhl\'{i}{\v{r}ov\'{a}, {J}. {V}{\'a}zquez, {F}. {W}ang,
  {J}.{D}. {W}atts, {P.} Yerg{\"u}n. {C}. {Z}hang, {X}. {Z}heng, and the
  integral packages {MOLECULE} ({J}. {A}lml{\"o}f and {P}.{R}. {T}aylor),
  {PROPS} ({P}.{R}. {T}aylor), {ABACUS} ({T}. {H}elgaker, {H}.{J}. {A}a.
  {J}ensen, {P}. {J}{\o}rgensen, and {J}. {O}lsen), and {ECP} routines by {A}.
  {V}. {M}itin and {C}. van {W}{\"u}llen. {F}or the current version, see
  \url{http://www.cfour.de}.}, {{CFOUR, Coupled-Cluster techniques for
  Computational Chemistry, a quantum-chemical program package}}}\BibitemShut
  {NoStop}%
\bibitem [{\citenamefont {Yamamoto}\ and\ \citenamefont
  {Tatewaki}(2015)}]{Yamamoto2015}%
  \BibitemOpen
  \bibfield  {author} {\bibinfo {author} {\bibfnamefont {S.}~\bibnamefont
  {Yamamoto}}\ and\ \bibinfo {author} {\bibfnamefont {H.}~\bibnamefont
  {Tatewaki}},\ }\bibfield  {title} {\enquote {\bibinfo {title} {Electronic
  spectra of {D}y{F} studied by four-component relativistic configuration
  interaction methods},}\ }\href {\doibase 10.1063/1.4913631} {\bibfield
  {journal} {\bibinfo  {journal} {J. Chem. Phys.}\ }\textbf {\bibinfo {volume}
  {142}},\ \bibinfo {pages} {094312} (\bibinfo {year} {2015})}\BibitemShut
  {NoStop}%
\bibitem [{\citenamefont {Kramers}(1930)}]{Kramers1930}%
  \BibitemOpen
  \bibfield  {author} {\bibinfo {author} {\bibfnamefont {H.~A.}\ \bibnamefont
  {Kramers}},\ }\bibfield  {title} {\enquote {\bibinfo {title} {Th\'{e}orie
  g\'{e}n\'{e}rale de la rotation paramagn\'{e}tique dans les cristaux},}\
  }\href {https://dwc.knaw.nl/DL/publications/PU00015981.pdf} {\bibfield
  {journal} {\bibinfo  {journal} {Proc. Amst. Acad.}\ }\textbf {\bibinfo
  {volume} {33}},\ \bibinfo {pages} {959--972} (\bibinfo {year}
  {1930})}\BibitemShut {NoStop}%
\bibitem [{\citenamefont {Tiesinga}\ \emph {et~al.}(2021)\citenamefont
  {Tiesinga}, \citenamefont {Mohr}, \citenamefont {Newell},\ and\ \citenamefont
  {Taylor}}]{CODATA2018}%
  \BibitemOpen
  \bibfield  {author} {\bibinfo {author} {\bibfnamefont {E.}~\bibnamefont
  {Tiesinga}}, \bibinfo {author} {\bibfnamefont {P.~J.}\ \bibnamefont {Mohr}},
  \bibinfo {author} {\bibfnamefont {D.~B.}\ \bibnamefont {Newell}}, \ and\
  \bibinfo {author} {\bibfnamefont {B.~N.}\ \bibnamefont {Taylor}},\ }\bibfield
   {title} {\enquote {\bibinfo {title} {{CODATA} recommended values of the
  fundamental physical constants: 2018},}\ }\href
  {https://link.aps.org/doi/10.1103/RevModPhys.93.025010} {\bibfield  {journal}
  {\bibinfo  {journal} {Rev. Mod. Phys.}\ }\textbf {\bibinfo {volume} {93}},\
  \bibinfo {pages} {025010} (\bibinfo {year} {2021})}\BibitemShut {NoStop}%
\bibitem [{\citenamefont {Kramida}\ \emph {et~al.}(2023)\citenamefont
  {Kramida}, \citenamefont {Ralchenko}, \citenamefont {Reader},\ and\
  \citenamefont {Team}}]{Kramida2023}%
  \BibitemOpen
  \bibfield  {author} {\bibinfo {author} {\bibfnamefont {A.}~\bibnamefont
  {Kramida}}, \bibinfo {author} {\bibfnamefont {Yu.}\ \bibnamefont
  {Ralchenko}}, \bibinfo {author} {\bibfnamefont {J.}~\bibnamefont {Reader}}, \
  and\ \bibinfo {author} {\bibfnamefont {NIST~ASD}\ \bibnamefont {Team}},\
  }\bibfield  {title} {\enquote {\bibinfo {title} {{NIST} {A}tomic {S}pectra
  {D}atabase (version 5.11), [{O}nline]},}\ }\href {\doibase
  https://doi.org/10.18434/T4W30F} {\  (\bibinfo {year} {2023}),\
  https://doi.org/10.18434/T4W30F},\ \bibinfo {note} {available:
  https://physics.nist.gov/asd. National Institute of Standards and Technology,
  Gaithersburg, MD.}\BibitemShut {Stop}%
\bibitem [{\citenamefont {Brink}\ and\ \citenamefont
  {Satchler}(1993)}]{Brink1993}%
  \BibitemOpen
  \bibfield  {author} {\bibinfo {author} {\bibfnamefont {D.~M.}\ \bibnamefont
  {Brink}}\ and\ \bibinfo {author} {\bibfnamefont {G.~R.}\ \bibnamefont
  {Satchler}},\ }\href@noop {} {\emph {\bibinfo {title} {Angular momentum}}},\
  \bibinfo {edition} {$3^{\rm rd}$}\ ed.\ (\bibinfo  {publisher} {Oxford
  University Press, Oxford},\ \bibinfo {year} {1993})\BibitemShut {NoStop}%
\bibitem [{\citenamefont {{\it et al}}(2016)}]{g16}%
  \BibitemOpen
  \bibfield  {author} {\bibinfo {author} {\bibfnamefont {M.~J.~Frisch}\
  \bibnamefont {{\it et al}}},\ }\href@noop {} {\enquote {\bibinfo {title}
  {Gaussian 16 {R}evision {B}.01},}\ } (\bibinfo {year} {2016}),\ \bibinfo
  {note} {{G}aussian Inc. Wallingford CT}\BibitemShut {NoStop}%
\bibitem [{\citenamefont {{\it et al}}(2015)}]{q-chem}%
  \BibitemOpen
  \bibfield  {author} {\bibinfo {author} {\bibfnamefont {Y.~Shao}\ \bibnamefont
  {{\it et al}}},\ }\bibfield  {title} {\enquote {\bibinfo {title} {Advances in
  molecular quantum chemistry contained in the {Q}-{C}hem 4 program package},}\
  }\href@noop {} {\bibfield  {journal} {\bibinfo  {journal} {Mol. Phys.}\
  }\textbf {\bibinfo {volume} {113}},\ \bibinfo {pages} {184--215} (\bibinfo
  {year} {2015})}\BibitemShut {NoStop}%
\bibitem [{\citenamefont {Chibotaru}\ and\ \citenamefont
  {Ungur}(2012)}]{Chibotaru2012}%
  \BibitemOpen
  \bibfield  {author} {\bibinfo {author} {\bibfnamefont {L.~F.}\ \bibnamefont
  {Chibotaru}}\ and\ \bibinfo {author} {\bibfnamefont {L.}~\bibnamefont
  {Ungur}},\ }\bibfield  {title} {\enquote {\bibinfo {title} {Ab initio
  calculation of anisotropic magnetic properties of complexes. {I}. {U}nique
  definition of pseudospin {H}amiltonians and their derivation},}\ }\href
  {\doibase 10.1063/1.4739763} {\bibfield  {journal} {\bibinfo  {journal} {J.
  Chem. Phys.}\ }\textbf {\bibinfo {volume} {137}},\ \bibinfo {pages} {064112}
  (\bibinfo {year} {2012})}\BibitemShut {NoStop}%
\bibitem [{\citenamefont {Malmqvist}\ \emph {et~al.}(1990)\citenamefont
  {Malmqvist}, \citenamefont {Rendell},\ and\ \citenamefont
  {Roos}}]{Malmqvist1990}%
  \BibitemOpen
  \bibfield  {author} {\bibinfo {author} {\bibfnamefont {P.-\AA.}\ \bibnamefont
  {Malmqvist}}, \bibinfo {author} {\bibfnamefont {A.}~\bibnamefont {Rendell}},
  \ and\ \bibinfo {author} {\bibfnamefont {B.~O.}\ \bibnamefont {Roos}},\
  }\bibfield  {title} {\enquote {\bibinfo {title} {The restricted active space
  self-consistent-field method, implemented with a split graph unitary group
  approach},}\ }\href {\doibase 10.1021/j100377a011} {\bibfield  {journal}
  {\bibinfo  {journal} {J. Phys. Chem.}\ }\textbf {\bibinfo {volume} {94}},\
  \bibinfo {pages} {5477--5482} (\bibinfo {year} {1990})}\BibitemShut {NoStop}%
\bibitem [{\citenamefont {Malmqvist}\ \emph {et~al.}(2002)\citenamefont
  {Malmqvist}, \citenamefont {Roos},\ and\ \citenamefont
  {Schimmelpfennig}}]{Malmqvist2002}%
  \BibitemOpen
  \bibfield  {author} {\bibinfo {author} {\bibfnamefont {P.-\AA.}\ \bibnamefont
  {Malmqvist}}, \bibinfo {author} {\bibfnamefont {B.~O.}\ \bibnamefont {Roos}},
  \ and\ \bibinfo {author} {\bibfnamefont {B.}~\bibnamefont
  {Schimmelpfennig}},\ }\bibfield  {title} {\enquote {\bibinfo {title} {The
  restricted active space ({RAS}) state interaction approach with spin–orbit
  coupling},}\ }\href {\doibase https://doi.org/10.1016/S0009-2614(02)00498-0}
  {\bibfield  {journal} {\bibinfo  {journal} {Chem. Phys. Lett.}\ }\textbf
  {\bibinfo {volume} {357}},\ \bibinfo {pages} {230--240} (\bibinfo {year}
  {2002})}\BibitemShut {NoStop}%
\bibitem [{\citenamefont {Roos}\ \emph {et~al.}(2008)\citenamefont {Roos},
  \citenamefont {Lindh}, \citenamefont {Malmqvist}, \citenamefont {Veryazov},
  \citenamefont {Widmark},\ and\ \citenamefont {Borin}}]{Roos2008}%
  \BibitemOpen
  \bibfield  {author} {\bibinfo {author} {\bibfnamefont {B.~O.}\ \bibnamefont
  {Roos}}, \bibinfo {author} {\bibfnamefont {R.}~\bibnamefont {Lindh}},
  \bibinfo {author} {\bibfnamefont {P.-\AA.}\ \bibnamefont {Malmqvist}},
  \bibinfo {author} {\bibfnamefont {V.}~\bibnamefont {Veryazov}}, \bibinfo
  {author} {\bibfnamefont {P.-O.}\ \bibnamefont {Widmark}}, \ and\ \bibinfo
  {author} {\bibfnamefont {A.~C.}\ \bibnamefont {Borin}},\ }\bibfield  {title}
  {\enquote {\bibinfo {title} {New relativistic atomic natural orbital basis
  sets for lanthanide atoms with applications to the {C}e diatom and
  {L}u{F}$_3$},}\ }\href {\doibase 10.1021/jp803213j} {\bibfield  {journal}
  {\bibinfo  {journal} {J. Phys. Chem. A}\ }\textbf {\bibinfo {volume} {112}},\
  \bibinfo {pages} {11431--11435} (\bibinfo {year} {2008})}\BibitemShut
  {NoStop}%
\bibitem [{\citenamefont {Matthews}\ \emph {et~al.}(2020)\citenamefont
  {Matthews}, \citenamefont {Cheng}, \citenamefont {Harding}, \citenamefont
  {Lipparini}, \citenamefont {Stopkowicz}, \citenamefont {Jagau}, \citenamefont
  {Szalay}, \citenamefont {Gauss},\ and\ \citenamefont
  {Stanton}}]{Matthews2020}%
  \BibitemOpen
  \bibfield  {author} {\bibinfo {author} {\bibfnamefont {D.~A.}\ \bibnamefont
  {Matthews}}, \bibinfo {author} {\bibfnamefont {L.}~\bibnamefont {Cheng}},
  \bibinfo {author} {\bibfnamefont {M.~E.}\ \bibnamefont {Harding}}, \bibinfo
  {author} {\bibfnamefont {F.}~\bibnamefont {Lipparini}}, \bibinfo {author}
  {\bibfnamefont {S.}~\bibnamefont {Stopkowicz}}, \bibinfo {author}
  {\bibfnamefont {T.-C.}\ \bibnamefont {Jagau}}, \bibinfo {author}
  {\bibfnamefont {P.~G.}\ \bibnamefont {Szalay}}, \bibinfo {author}
  {\bibfnamefont {J.}~\bibnamefont {Gauss}}, \ and\ \bibinfo {author}
  {\bibfnamefont {J.~F.}\ \bibnamefont {Stanton}},\ }\bibfield  {title}
  {\enquote {\bibinfo {title} {Coupled-cluster techniques for computational
  chemistry: The {CFOUR} program package},}\ }\href {\doibase
  10.1063/5.0004837} {\bibfield  {journal} {\bibinfo  {journal} {J. Chem.
  Phys.}\ }\textbf {\bibinfo {volume} {152}},\ \bibinfo {pages} {214108}
  (\bibinfo {year} {2020})}\BibitemShut {NoStop}%
\bibitem [{\citenamefont {Purvis}\ and\ \citenamefont
  {Bartlett}(1982)}]{Purvis1982}%
  \BibitemOpen
  \bibfield  {author} {\bibinfo {author} {\bibfnamefont {G.~D.~III}\
  \bibnamefont {Purvis}}\ and\ \bibinfo {author} {\bibfnamefont {R.~J.}\
  \bibnamefont {Bartlett}},\ }\bibfield  {title} {\enquote {\bibinfo {title} {A
  full coupled‐cluster singles and doubles model: The inclusion of
  disconnected triples},}\ }\href {\doibase 10.1063/1.443164} {\bibfield
  {journal} {\bibinfo  {journal} {J. Chem. Phys.}\ }\textbf {\bibinfo {volume}
  {76}},\ \bibinfo {pages} {1910--1918} (\bibinfo {year} {1982})}\BibitemShut
  {NoStop}%
\bibitem [{\citenamefont {Raghavachari}\ \emph {et~al.}(1989)\citenamefont
  {Raghavachari}, \citenamefont {Trucks}, \citenamefont {Pople},\ and\
  \citenamefont {Head-Gordon}}]{Raghavachari1989}%
  \BibitemOpen
  \bibfield  {author} {\bibinfo {author} {\bibfnamefont {K.}~\bibnamefont
  {Raghavachari}}, \bibinfo {author} {\bibfnamefont {G.~W.}\ \bibnamefont
  {Trucks}}, \bibinfo {author} {\bibfnamefont {J.~A.}\ \bibnamefont {Pople}}, \
  and\ \bibinfo {author} {\bibfnamefont {M.}~\bibnamefont {Head-Gordon}},\
  }\bibfield  {title} {\enquote {\bibinfo {title} {A fifth-order perturbation
  comparison of electron correlation theories},}\ }\href {\doibase
  https://doi.org/10.1016/S0009-2614(89)87395-6} {\bibfield  {journal}
  {\bibinfo  {journal} {Chem. Phys. Lett.}\ }\textbf {\bibinfo {volume}
  {157}},\ \bibinfo {pages} {479--483} (\bibinfo {year} {1989})}\BibitemShut
  {NoStop}%
\bibitem [{\citenamefont {Stanton}\ and\ \citenamefont
  {Bartlett}(1993)}]{Stanton1993}%
  \BibitemOpen
  \bibfield  {author} {\bibinfo {author} {\bibfnamefont {J.~F.}\ \bibnamefont
  {Stanton}}\ and\ \bibinfo {author} {\bibfnamefont {R.~J.}\ \bibnamefont
  {Bartlett}},\ }\bibfield  {title} {\enquote {\bibinfo {title} {The equation
  of motion coupled‐cluster method. a systematic biorthogonal approach to
  molecular excitation energies, transition probabilities, and excited state
  properties},}\ }\href {\doibase 10.1063/1.464746} {\bibfield  {journal}
  {\bibinfo  {journal} {J. Chem. Phys.}\ }\textbf {\bibinfo {volume} {98}},\
  \bibinfo {pages} {7029--7039} (\bibinfo {year} {1993})}\BibitemShut {NoStop}%
\bibitem [{\citenamefont {Liu}\ and\ \citenamefont {Cheng}(2018)}]{Liu2018}%
  \BibitemOpen
  \bibfield  {author} {\bibinfo {author} {\bibfnamefont {J.}~\bibnamefont
  {Liu}}\ and\ \bibinfo {author} {\bibfnamefont {L.}~\bibnamefont {Cheng}},\
  }\bibfield  {title} {\enquote {\bibinfo {title} {An atomic mean-field
  spin-orbit approach within exact two-component theory for a non-perturbative
  treatment of spin-orbit coupling},}\ }\href {\doibase 10.1063/1.5023750}
  {\bibfield  {journal} {\bibinfo  {journal} {J. Chem. Phys.}\ }\textbf
  {\bibinfo {volume} {148}},\ \bibinfo {pages} {144108} (\bibinfo {year}
  {2018})}\BibitemShut {NoStop}%
\bibitem [{\citenamefont {Zhang}\ and\ \citenamefont
  {Cheng}(2022)}]{Zhang2022}%
  \BibitemOpen
  \bibfield  {author} {\bibinfo {author} {\bibfnamefont {C.}~\bibnamefont
  {Zhang}}\ and\ \bibinfo {author} {\bibfnamefont {L.}~\bibnamefont {Cheng}},\
  }\bibfield  {title} {\enquote {\bibinfo {title} {Atomic mean-field approach
  within exact two-component theory based on the {D}irac–{C}oulomb–{B}reit
  hamiltonian},}\ }\href {\doibase 10.1021/acs.jpca.2c02181} {\bibfield
  {journal} {\bibinfo  {journal} {J. Phys. Chem. A}\ }\textbf {\bibinfo
  {volume} {126}},\ \bibinfo {pages} {4537--4553} (\bibinfo {year}
  {2022})}\BibitemShut {NoStop}%
\bibitem [{\citenamefont {Dunning}(1989)}]{Dunning1989}%
  \BibitemOpen
  \bibfield  {author} {\bibinfo {author} {\bibfnamefont {T.~H.~Jr.}\
  \bibnamefont {Dunning}},\ }\bibfield  {title} {\enquote {\bibinfo {title}
  {Gaussian basis sets for use in correlated molecular calculations. {I}. the
  atoms boron through neon and hydrogen},}\ }\href {\doibase 10.1063/1.456153}
  {\bibfield  {journal} {\bibinfo  {journal} {J. Chem. Phys.}\ }\textbf
  {\bibinfo {volume} {90}},\ \bibinfo {pages} {1007--1023} (\bibinfo {year}
  {1989})}\BibitemShut {NoStop}%
\end{thebibliography}%

\end{document}